\documentclass[a4paper,11pt]{article}
\pdfoutput=1 % if your are submitting a pdflatex (i.e. if you have
\usepackage{graphicx} % needed for figures
\usepackage{dcolumn} % needed for some tables
\usepackage{bm} % for math
\usepackage{amssymb} % for math
\usepackage[latin1]{inputenc}
\usepackage[english]{babel}
\usepackage{shapepar}
\usepackage{jcappub}
\usepackage[T1]{fontenc} % if needed
\usepackage{amsmath}
\usepackage{amsthm}
\usepackage{array}
\usepackage{fancybox}
\usepackage{multirow}
\usepackage{appendix}
\usepackage{epsfig}
\usepackage{amsmath}
\usepackage{amsthm}
\usepackage{array}
\usepackage{fancybox}
\usepackage{multirow}
\usepackage{appendix}
\usepackage{epsfig}
\usepackage{cases}
\usepackage{xcolor}
\usepackage{enumitem}
\newcommand{\D}{\partial}
\title{\boldmath Disformal couplings in a $\Lambda$CDM background cosmology}
\author[1]{Avishek Dusoye}
\author[1]{, \'Alvaro de la Cruz-Dombriz}
\author[1]{, Peter Dunsby}
\author[2]{, Nelson J. Nunes.}
\affiliation[1]{Cosmology and Gravity Group, Department of Mathematics and Applied Mathematics,\\ University of Cape Town, Rondebosch 7701, Cape Town, South Africa.}
\affiliation[2]{Instituto de Astrof\'isica e Ci\^encias do Espa\c{c}o,
Faculdade de Ci\^encias da Universidade de Lisboa, Campo Grande, PT1749-016 
Lisboa, Portugal.}
\emailAdd{avsdus001@myuct.ac.za}
\emailAdd{alvaro.delacruzdombriz@uct.ac.za}
\emailAdd{peter.dunsby@uct.ac.za}
\emailAdd{njnunes@fc.ul.pt}
\abstract{The coupled quintessence model with disformal couplings is treated here to mimic the $\Lambda$CDM background. Using this approach, the quintessence potential does not have to be specified. The model considers a generic fluid coupled to the quintessence, which is specified to be either dark matter or a relativistic fluid. The background consists of a cosmological constant and another uncoupled generic fluid, to cater for three  studied scenarios. The cosmological dynamics is investigated for the coupled quintessence model, whose disformal couplings depend on the equation of state of both generic fluids. The scenario, whereby both generic fluids are dark matter, was further analysed for an expansion history of the mimicking model. The analysis confirms that the mass scale of the quintessence influences the disformal characteristics of the dynamical system, which is portrayed by the evolution of an effective conformal coupling.}

\begin{document}
\maketitle
\flushbottom 

\newpage
\section{Introduction}
%%%%%%%%%%%%%%%%%%%%%%%%%%%%%%%%%%%%%%%
The Cosmological Constant problem remains the largest unsolved mystery in modern cosmology, even though it has been known for more than a decade now that a Dark Energy (DE) component  is required to fit the observational data \cite{Riess1998,Sullivan2011}. Planck's latest results in 2018, combined with Type Ia Supernovae, measures an equation of state (EOS) parameter for DE of $w_{\text{DE}}= -1.03\pm0.03$ \cite{Plank2018}. Although the nature of DE lacks an exact theoretical explanation, the $\Lambda$CDM remains one of the most successful cosmological models so far from an observational perspective \cite{Peacock1999}. According to the $\Lambda$CDM  paradigm, $\Lambda$ and DM consist of about 68\% and 27\% respectively of the content of the Universe and there is compelling observational evidence pointing to their existence \cite{Plank2018,Allen2011,Clowe2003,Corbelli1999}. This strongly motivates our main assumption in this work in particular, that the cosmological expansion history is well described by the $\Lambda$CDM model.

A range of theoretical possibilities have been proposed beyond $\Lambda$CDM. The first category makes use of canonical scalar fields \cite{Thorsrud2019} or effective fluids \cite{Nunes2000}. Whether a scalar field or effective fluids are considered, they both have the possibility of interacting with each other. The second category is modified gravity \cite{Clifton2012}, which involves the extensions to GR such as considering a $f(R)$ correction to the gravitational action and other scalar-tensor theories.

In this article, we focus on the cosmological dynamics of the quintessence model, under a number of assumptions (discussed below), which lead to several new perspectives. It is normally assumed that the dark sector can only be observed through gravitational effects because this form of mass-energy does not interact directly with ordinary (baryonic) matter and light.  Hence, the possibility that quintessence interacts with DM is an interesting avenue to investigate \cite{Wang2016}. Such interactions (referred to as coupled quintessence) were first introduced in \cite{Amendola1999,Holden1999}. In what follows, we first assume (for the sake of simplicity) that the quintessence is allowed to interact with a single generic and effective fluid and we investigate the novel phenomenology that arises due to these interactions. The second assumption concerns the geometry in which the quintessence exists. In fact, the theory of gravitation might actually require two geometries: (i) a gravitational geometry - describing the curvature on the fabric of space-time, and (ii) a physical geometry - which describes how the matter fields propagate. There exist many such examples \cite{BransDicke1962,Dirac1973,HOROWITZ1991}. If $g_{ab}$ is the metric of the gravitational geometry, and $\tilde{g}_{ab}$ is the metric of the physical geometry, then both metrics could be related to each other by the conformal transformation below, where $C(\phi)$ is the conformal function and usually depends of the quintessence field $\phi$:
\begin{equation}\label{E1}
\tilde{g}_{ab} = C(\phi)g_{ab}\;\;.
\end{equation}
The effect of the conformal transformation can be understood as rescaling the length of original metric $g_{ab}$ \cite{Clifton2012}. Hence we have  $C(\phi)=1$ in the case of GR, which is based on single-geometric description. Any dually geometric theory of gravity i.e., $C(\phi)\neq1$, will break the strong equivalence principle \cite{Bekenstein1992}. However, in such theories of gravity, the relation between the physical metric to gravitational metric is still constrained by the weak equivalence principle and causality \cite{Bekenstein1992}. Moreover, in dually geometric gravitation, the only way to maintain the principle of general covariance is the addition of new fields  and hence the use of the quintessence seems convenient \cite{Bekenstein1992}. As shown by \cite{Bekenstein1992}, the disformal relation is obtained by extending the conformal transformation with an extra scaled kinetic term of the scalar field:
\begin{equation}\label{E2}
\tilde{g}_{ab} = C(\phi)g_{ab} + D(\phi)\D_{a}\phi\D_{b}\phi\;\;,
\end{equation}
where $D(\phi)$ is the disformal coefficient. The effect of a disformal transformation can be understood as the distortion of angles and lengths in the original metric  $g_{ab}$ due to compression, which occurs in the direction of the gradient of the field \cite{Bekenstein1992,Teixeira2019}. The occurence of the disformal relation happens in many theories such as Horndeski-type scalar tensor theory \cite{Zuma2013}\cite{Bettoni2013}, non-linear massive gravity theory \cite{deRham2010,Do2016}, brane cosmology with higher dimensions \cite{Koivisto2013} and this framework has attracted considerable attention lately \cite{Teixeira2019,Gannouji2018,Brax2019,Xiao2018}.

 The popular $\Lambda$CDM cosmology is known to bring about tensions between the relatively high level of clustering found in cosmic microwave background experiments and the smaller one obtained from large-scale observations in the late Universe. A way to alleviate this issue is to consider a scalar field dark energy component conformally coupled to dark matter maintaining a $\Lambda$CDM background cosmology \cite{Barros2018,Barros2019}. In this article we extend these studies to allow in addition disformal couplings, which we explore by performing a dynamical system analysis \cite{Landim2016,Shahalam2015, Boehmer2015}.
 
The article is organised in the following order; Section I motivates the premises upon which our cosmological model is built. In Section II, the Lagrangian description of the coupled quintessence model is provided under disformal transformation. Section III presents how the coupled quintessence model mimics the $\Lambda$CDM background, which explains the reasoning behind not having to specify the quintessence potential. In Section IV, the fluid continuity equations are described as a dynamical system using dimensionless dynamical variables and the bounds of a compact phase space are defined to represent the behaviour of the dynamical system. In Section V, a brief description of the three studied scenarios is given, as well as the suggested approach used to analyse the dynamical system. Section VI entails the analysis of the conformal couplings in the three studied scenarios, which includes the conformal equations, the fixed points, and 2D phase portrait. Sections VII and VIII present the analysis of the disformal couplings in the three studied scenarios, which includes the disformal equations, the additional disformal fixed points, the topological features and the investigation of the  trajectories on a 3D phase portrait. Finally in Section IX, we shall present our remarks about the expansion history of the coupled quintessence of Secnario I (where both generic fluids are DM) from the early universe to present, and (ii) the effect of the mass scale on the interaction.

%\newpage
\section{Constructing the model}
%%%%%%%%%%%%%%%%%%%%%%%%%%%%%%%%%%%%%%%%%%%%%%%%%%%%%%%%
The well known theoretical idea of quintessence is to have a scalar field $\phi$, which can substitute the cosmological constant $\Lambda$. Therefore, the Lagrangian function  for the quintessence $\mathcal{L}_{\phi}$, which depends on the gravitational metric $g^{ab}$, should consist of a kinetic and a potential function:
\begin{equation} \label{E3}
\mathcal{L}_{\phi} = -\dfrac{1}{2}g^{ab}\partial_{a}\phi\partial_{b}\phi - V(\phi).
\end{equation}
The Lagrangian for matter $\mathcal{L}_{m}$, on the other hand, depends on the physical metric $\tilde{g}^{ab}$, which defines the geodesic upon which the matter fields $\psi$ propagate. If the Einstein-Hilbert action is extended to include the quintessence, the new action in the Einstein frame becomes:
\begin{equation} \label{E4}
S =  \int \! \left( \dfrac{1}{2 \kappa^2}R \; +  \;\mathcal{L}_{\phi}(g_{ab} ,\phi)\;  + \;\mathcal{L}_{m}(\tilde{g}_{ab} ,\psi)  \right) \,\sqrt{-g} \;\rm{d}^{4}x \; ,
\end{equation}
where the first term is the standard GR gravitational action, with Ricci scalar $R$ and  $\kappa^2 \equiv 8\pi G = M_{\rm{pl}}^{-2}$. The Planck's mass is taken as $ M_{\rm{pl}} = 2.4 \times 10^{18} \; \rm{GeV}$ with the speed of light and reduced Planck's constant to be assumed as unity $(\hbar = c  = 1)$. The metric variation of \eqref{E4} with respect to the metric $g^{ab}$ results in the Einstein field equations in the Einstein frame:
\begin{equation} \label{E5}
G_{ab} \equiv R_{ab} - \dfrac{1}{2}g_{ab}R = \kappa^2 T_{ab}^{(tot)} =\;   \kappa^2 ( T_{ab}^{\phi} + T_{ab}) \;,
\end{equation}
where $G_{ab}$ and $R_{ab}$ are the Einstein and Ricci tensors respectively. The field equations can be re-arranged to express the quintessence and a coupled fluid as two separate source terms, $T_{ab}^{\phi}$ and $T_{ab} $ respectively. The energy-momentum tensors $T_{ab}^{\phi}$ and  $T_{ab}$ are usually defined as:
\begin{equation}\label{E6}
T_{ab}^{\phi} = -\dfrac{2}{\sqrt{-g}}\dfrac{\delta (\sqrt{-g}\mathcal{L}_{\phi})}{\delta g^{ab}} \hspace{0.25cm} \text{and}\hspace{0.25cm} T_{ab} = -\dfrac{2}{\sqrt{-g}}\dfrac{\delta (\sqrt{-g}\mathcal{L}_{m})}{\delta g^{ab}} \hspace{0.25cm}\text{such that}\hspace{0.25cm}
 \tilde{T}_{ab} = \dfrac{\sqrt{-g}}{\sqrt{-\tilde{g}}}\dfrac{\delta g^{ab}}{\delta \tilde{g}^{ab}} T_{ab}\;,
 \end{equation}
where the energy-momentum tensor of the disformal frame (denoted by $\tilde{T}_{ab}$) can still be related to that of the Einstein frame under the transformation $\tilde{T}_{ab} \rightarrow  T_{ab}$. The Jacobian of the disformal relation  \eqref{E2} provides how the determinants of the metrics are related.
\begin{equation}\label{E7}
J = \dfrac{\sqrt{-g}}{\sqrt{-\tilde{g}}} =C^{2}(\phi)\sqrt{1+\dfrac{D}{C}g^{ab}\D_{a}\phi\D_{b}\phi }\;.
\end{equation}
Moreover, the energy-momentum tensor of any perfect fluid reduces to:
\begin{equation}
T^{ab}_{l} = (\rho_l + P_l)u^{a}u^{b} + P_l g^{ab}\;,
\end{equation}
Where $\rho_l $ and $P_l$ are the energy density and pressure respectively for an $l^{th}$ fluid moving with four velocity $u^{a}$ with respect to comoving observer. In our case, the metric $g_{ab}$ is now specified to be spatially flat Friedmann-Lema\^{i}tre-Robertson-Walker (FLRW), which inherently assumes a homogeneous and isotropic universe:
\begin{equation}  \label{E16}
\rm{d}s^2 = g_{ab}\rm{d}x^{a}\rm{d}x^{b} = -\rm{d}t^2 + a^2(t)\;\delta_{\textit{ij}}\rm{d}x^{\textit{i}}\rm{d}x^{\textit{j}} \;,
\end{equation}
where $a(t)$ is the scale factor, $t$ is the cosmic time and $i,j= 1,2,3$ denote the spatial coordinates. After contracting the Ricci tensor $R_{ab}$ in the Einstein equations \eqref{E5}, the spatial Ricci scalar $^{(3)}R$ hence gives us a general form of the Friedmann equation, which then simplifies into:
\begin{equation} \label{E17}
  H^{2}  = \dfrac{\kappa^2}{3}(\rho_{\phi} + \rho_c)\;,
 \hspace*{1.0 cm} \text{and} \hspace*{1.0 cm} 
  \dot{H} = -\dfrac{1}{2}\kappa^2\left[ (\rho_{\phi} + P_{\phi}) + (1+w_c)\rho_c\right] \;,
\end{equation}
where $H$ is the Hubble parameter $H \equiv \dot{a}/a$. The energy density and the pressure for the quintessence in the Einstein frame can be computed from the components of $T^{ab}_{\phi}$ to be:
\begin{equation} \label{E9}
\rho_{\phi} \equiv \dfrac{1}{2}\dot{\phi}^{2} + V(\phi) \hspace{0.5cm}\text{and}\hspace{0.5cm} P_{\phi} \equiv \dfrac{1}{2}\dot{\phi}^{2} - V(\phi)\;, \hspace{0.5cm}\text{with}\hspace{0.5cm} w_{\phi} = \dfrac{P_{\phi} }{\rho_{\phi}} \;.
\end{equation}
As mentioned earlier, the first premise is that the quintessence is allowed to interact with a single generic and effective perfect fluid with EOS parameter $w_{\rm{c}}$. Assuming $T^{ab}$ is a perfect fluid, the energy density, the pressure and the EOS parameter for the coupled fluid in Einstein frame can be mapped into the disformal frame (See Appendix A.4 of \cite{vandeBruck2015c}):
\begin{equation}\label{E8}
\tilde{\rho_c}=\dfrac{\rho_c}{C^{2}}\sqrt{1+\dfrac{D}{C}g^{ab}\D_{a}\phi\D_{b}\phi }\;,\hspace{0.2cm}
\tilde{P}_{c}=\dfrac{P_{c}}{C^{2}\sqrt{1+\dfrac{D}{C}g^{ab}\D_{a}\phi\D_{b}\phi }}\;,\hspace{0.2cm}
\tilde{w}\equiv \dfrac{\tilde{P}_{c}}{\tilde{\rho}_{c}}=\dfrac{w_{c}}{\left(1+\dfrac{D}{C}g^{ab}\D_{a}\phi\D_{b}\phi \right)}\;.
\end{equation}
 If we further assume the quintessence to be isotropic (i.e. $\phi =\phi(t)$ is only a function of time), then in the Einstein frame, the EOS parameter for the coupled fluid and its time-derivative are as below, where the dot means derivative with respect to cosmic time:
\begin{eqnarray}\label{E10}
w_{\rm{c}}= \left(1-\dfrac{D}{C}\dot{\phi}^2 \right)\tilde{w}
\hspace{1.0cm}\text{and}\hspace{1.0cm}
\dot{w}_{\rm{c}}= - \tilde{w}\dot{\phi}\dfrac{D}{C}\left[ \left(\dfrac{D_{,\phi}}{D}-\dfrac{C_{,\phi}}{C}\right)\dot{\phi}^2 + 2\ddot{\phi} \right] \;.
\end{eqnarray}
The variation of the action \eqref{E4} with respect to the quintessence $\phi$ results in the modified Klein-Gordon equation, with $\square \equiv g^{ab}\nabla_{a}\nabla_{b} $, being the D'Alembertian operator \cite{Carsten2016a}:
\begin{equation}\label{E11}
\square\phi = V_{,\phi} - Q(\phi) \;,
\end{equation}
where $V_{,\phi}$ denotes the derivative of $V(\phi)$ with respect to $\phi$, and the interaction term for the coupling between the quintessence and a single fluid is given by \cite{Carsten2016a}:  
\begin{equation}\label{E13}
Q(\phi) = \dfrac{C_{,\phi}}{2C}T + \dfrac{D_{,\phi}}{2C}T^{ab}\nabla_{a}\phi\nabla_{b}\phi - \nabla_{a}\left( \dfrac{D}{C}T^{ab} \nabla_{b}\phi \right) \;,
\end{equation}
where $C_{,\phi}$ and $D_{,\phi}$ are the  derivatives of conformal and disformal coefficients $C(\phi)$ and $D(\phi)$ respectively. From the expression provided in \eqref{E6}, one can derive the following conservation equations for the energy-momentum tensors $T_{ab}$ and $T_{ab}^{\phi}$:
\begin{eqnarray} \label{E12}
\nabla^{a}T_{ab}^{\phi}= -Q\D _{b} \phi  \;,  \hspace{0.5cm} \text{and} \hspace{0.5cm}
\nabla^{a}T_{ab}= +Q \D _{b} \phi    \;,                                                          
\end{eqnarray} 
such that it follows that there is a flow of energy and momentum between two coupled fluids in the model, which is dictated by the interaction term $Q(\phi)$. The total energy-momentum tensor is conserved although their individual components are not separately conserved. By projecting the conservation equations \eqref{E12} along $u_{a}$, we obtain the continuity equations for each fluid:
 \begin{eqnarray}
 \ddot{\phi} +3H\dot{\phi} + V_{,\phi} &=& +Q \;\;,           \label{E14} \\
 \dot{\rho}_{\rm{c}} + 3H\rho_{\rm{c}}(1+w_{\rm{c}}) &=& -Q\dot{\phi} \;.                 \label{E15}
 \end{eqnarray}
The Einstein field equations \eqref{E5}, the modified Klein-Gordon equation \eqref{E11} and the disformal interaction term $Q(\phi)$ in equation \eqref{E13} may be applied to general cosmological models. 

%%%%%%%%%%%%%%%%%%%%%%%%%%%%%%%%%%%%%%%%%%%%%%
\section{Mimicking the $\Lambda$CDM background}
%%%%%%%%%%%%%%%%%%%%%%%%%%%%%%%%%%%%%%%%%%%%%%
 In the following, we consider a cosmological background, which has only two fluids, i.e., the cosmological constant $\Lambda$ and an uncoupled generic fluid $\rho_{u}$, whose EOS parameter is $w_{u}$. In the exact $\Lambda$CDM background,  the uncoupled generic fluid is simply the cold dark matter $\rho_{\rm{cdm}}$ but we rather treat this background with uncoupled generic fluid to investigate other interesting scenarios (See Sections VI, VII, VIII below). Assuming the FLRW metric  \eqref{E16}, the Friedmann and Raychaudhuri equations for this  background are:
\begin{equation} \label{E18}
  H^{2}_{\Lambda \rm{CDM}}  = \dfrac{\kappa^2}{3}(\rho_{\Lambda} + \rho_u)\;,
 \hspace*{1.0 cm} \text{and} \hspace*{1.0 cm}
  \dot{H}_{\Lambda \rm{CDM}}  = -\dfrac{1}{2}\kappa^2\left[ (\rho_{\Lambda} + P_{\Lambda}) + (1+w_u)\rho_u\right]\;. 
\end{equation}
%%%%%%%%%%%%%%%%%%%%%%%%%%%%%%%%%%%%%%%%%%%%%%%%%%%%%%%%
 On the other hand, the quintessence model, constructed in Section II, also has two fluids i.e, the quintessence $\phi$ and the coupled fluid $\rho_{c}$, whose EOS parameter is $w_{c}$. As mentioned in the Introduction, the novelty of this work is to extend the cosmological background of the coupled quintessence, mimicking $\Lambda$CDM background  (See \cite{ Barros2018}), to include a disformal coupling. Hence after equating their Friedmann equations  \eqref{E17} and \eqref{E18} from the two cosmological backgrounds, we can express the energy density of the quintessence in terms of the energy densities for the cosmological constant and the coupled fluid, and that of uncoupled fluid:
\begin{eqnarray} \label{E19}
H^{2} =   \dfrac{\kappa^2}{3}(\rho_{\phi} + \rho_c)& = &  \dfrac{\kappa^2}{3}(\rho_{\Lambda} + \rho_u)  = H^{2}_{\Lambda \rm{CDM}}\;,\\ \label{E20}
\rho_{\phi}&  = &  \rho_{\Lambda} +\rho_{u} - \rho_{c} \;.
\end{eqnarray} 
By equating their Raychaudhuri equations from the two cosmological backgrounds $\dot{H} = \dot{H}_{\Lambda CDM}$, we can express the pressure of the quintessence in terms of the pressure  for the cosmological constant and the pressure of the uncoupled fluid, and that of coupled fluid:
\begin{eqnarray} \label{E21}
 \dot{H} = -\dfrac{1}{2}\kappa^2\left[ (\rho_{\phi} + P_{\phi}) + (1+w_c)\rho_c\right] & = &  -\dfrac{1}{2}\kappa^2\left[ (\rho_{\Lambda} + P_{\Lambda}) + (1+w_u)\rho_u\right]   =   \dot{H}_{\Lambda \rm{CDM}} \\ \label{E22}
P_{\phi} & = &  -\rho_{\Lambda} + w_{u}\rho_{u} - w_{c}\rho_{c} \;.
\end{eqnarray} 
One can therefore write $(\rho_{\phi} + P_{\phi})$ using equations \eqref{E20} and \eqref{E22} to find an expression for $\rho_{u}$:
\begin{eqnarray}\label{E23}
\rho_{u} & = & \dfrac{\dot{\phi}^{2}}{1+ w_{u}}+ \rho_{c}\;\frac{1+w_{c}}{1+ w_{u}}\;.
\end{eqnarray}
Then after plugging $\rho_{u}$ back into equation \eqref{E20} using \eqref{E23}, the expression for $\rho_{\phi}$ in terms of quintessence $\phi$ is obtained:
\begin{eqnarray}\label{E26}
\rho_{\phi} &=&  \rho_{\Lambda}  +  \dfrac{\dot{\phi}^{2}}{1+ w_{u}}+ \rho_{c}\;\frac{w_{c}-w_{u}}{1+ w_{u}}\;. 
\end{eqnarray}
Substituting equation \eqref{E26}  and $ \dot{\phi}^{2} =\rho_{\phi} + P_{\phi}$ into the Friedmann and  the Raychaudhuri equations \eqref{E17} respectively, we obtain the following form:
\begin{equation}
  H^{2}  = \dfrac{\kappa^2}{3}\left[ \dfrac{\dot{\phi}^{2}}{1+ w_{u}} + \rho_{\Lambda} +\rho_{c}\left(\frac{1+w_{c}}{1+ w_{u}}\right) \right]\;,
  \hspace*{1.0 cm} 
  \dot{H} = -\dfrac{1}{2}\kappa^2\left[\; \dot{\phi}^{2} +\rho_{c} (1+w_c)\;\right]\;.     \label{E27}
\end{equation}
From equation \eqref{E27}, when $\dot{\phi}= 0$, it follows that the coupled system is exactly mimicking $\Lambda$CDM with an uncoupled fluid. This assumption of mimicking an uncoupled setup brings the convenience of not having to define a potential. The potential function of the quintessence and its time-derivative is obtained by inserting relation \eqref{E26} into $V(\phi) = \rho_{\phi} - \frac{1}{2}\dot{\phi}^{2}$ and are given below:
\begin{eqnarray} \label{E28}
V(\phi) & = & \rho_{\Lambda}  + \dfrac{\;\dot{\phi}^{2}}{2}\left(\dfrac{1- w_{u}}{1+ w_{u}}\right) + \rho_{c}\frac{w_{c}-w_{u}}{1+ w_{u}} \;, \\ \label{E29}
%%%%%%%%%%%%%%%%%%%%
\dot{V}= V_{,\phi}\dot{\phi} & = & \dot{\phi}\ddot{\phi}\;\dfrac{1- w_{u}}{1+ w_{u}} + \dot{\rho}_{c}\frac{w_{c}-w_{u}}{1+ w_{u}} + \rho_{c}\dot{w_{c}}\;.
\end{eqnarray}
Moreover, the EOS parameter for an uncoupled fluid is the same constant in both Einstein frame and the disformal frame, i.e. $w_{u} = \tilde{w}_{u} = \text{constant} \Rightarrow \dot{w}_{\textit{u}} = 0$, whereas for the coupled fluid, the EOS parameter $w_{c}$ and its time-derivative $\dot{w}_{c}$ are given by \eqref{E10}. Therefore plugging \eqref{E10} and the continuity equation \eqref{E15} into \eqref{E29} and after simplifying, we obtain the derivative of the potential $V_{,\phi}$:
\begin{eqnarray}\label{E30}
V_{,\phi} & = & \dfrac{\ddot{\phi}}{1+ w_{u}}\left(1- w_{u} - 2\tilde{w}\rho_{c}\dfrac{D}{C}\right)\; + \; \dfrac{\tilde{w}\rho_{c}}{1+ w_{u}}\left(\dfrac{D_{,\phi}}{D}-\dfrac{C_{,\phi}}{C}\right)\dfrac{D}{C}\dot{\phi}^2  \nonumber \\
&& - \left(\dfrac{3H\rho_{c}}{\dot{\phi}}\right)\dfrac{(w_{c}-w_{u})(1+ w_{c})}{1+ w_{u}}\; - \dfrac{w_{c}-w_{u}}{1+ w_{u}}Q(\phi)\;.
\end{eqnarray}
This  $V_{,\phi} $ in equation  \eqref{E30} can be plugged into the continuity equation \eqref{E14} to find an expression for $\ddot{\phi}$, which is useful when computing the interaction term $Q(\phi)$:
\begin{small}
\begin{eqnarray} \label{E31}
\hspace{-1.0cm} \ddot{\phi} = \dfrac{(1+ w_{c})Q(\phi) - 3H\dot{\phi}(1+ w_{u}) + \tilde{w}\rho_{c}\left(\dfrac{D}{C}\right)\left(\dfrac{D_{,\phi}}{D}-\dfrac{C_{,\phi}}{C}\right)\dot{\phi}^2 + \left(\dfrac{3H\rho_{c}}{\dot{\phi}}\right)(w_{c}-w_{u})(1+ w_{c})}{\left(2-2\tilde{w}\rho_{c}\dfrac{D}{C}\right)}.
\end{eqnarray}
\end{small}
%%%%%%%%%%%%%%%%%%%%%%%%%%%%%%%%%%%%%%%%%%%%%%%%%%%%%%%%
Lastly, the interaction term $Q(\phi)$ for a single generic coupled fluid can be computed by evaluating the covariant derivative of time components for the energy-momentum tensor in equation \eqref{E13}: 
\begin{small}
\begin{eqnarray} \label{E32}
Q(\phi) =-\dfrac{\rho_c}{2}\dfrac{C_{,\phi}}{C}(1 -3 w_c) -\rho_{c}\dfrac{D}{C}\left(\dfrac{D_{,\phi}}{D}-\dfrac{C_{,\phi}}{C}\right)\dot{\phi}^2 
 + \dfrac{D}{C}Q\dot{\phi}^2+3\dfrac{D}{C}w_{c}\rho_{c}H\dot{\phi}- \rho_{c}\ddot{\phi} \dfrac{D}{C}\;.
\end{eqnarray}
 \end{small}
After plugging \eqref{E31} into the last term of equation \eqref{E32}, one can obtain this form for the interaction term: 
\begin{eqnarray} \label{E33}
Q(\phi) &=&\dfrac{-\rho_{c}\left(2-2\tilde{w}\dfrac{D}{C}\right)\left[ \dfrac{C_{,\phi}}{2C}(1 -3 w_c) + \dfrac{D}{C}\left(\dfrac{D_{,\phi}}{2D}-\dfrac{C_{,\phi}}{C}\right)\dot{\phi}^2   - 3Hw_{c}\dfrac{D}{C}\dot{\phi}  \right]}{\left[\left( 1- \dfrac{D}{C}\dot{\phi}^2\right)\left(2-2\tilde{w}\rho_{c}\dfrac{D}{C}\right) + \rho_{c}(1+ w_{c})\left( \dfrac{D}{C}\right)\right]} +  \\
%%%%%%%%%%%%%%%%%%%%%%%%%%%%%%%%
 &&  \frac{\dfrac{D\rho_{c}}{C}\left[ 3H\dot{\phi}(1+ w_{u}) - \tilde{w}\rho_{c}\left(\dfrac{D}{C}\right)\left(\dfrac{D_{,\phi}}{D}-\dfrac{C_{,\phi}}{C}\right)\dot{\phi}^2 - \left(\dfrac{3H\rho_{c}}{\dot{\phi}}\right)(w_{c}-w_{u})(1+ w_{c})\right] }{\left[\left( 1- \dfrac{D}{C}\dot{\phi}^2\right)\left(2-2\tilde{w}\rho_{c}\dfrac{D}{C}\right) + \rho_{c}(1+ w_{c})\left( \dfrac{D}{C}\right)\right]}\;. \nonumber
%%%%%%%%%%%%%%%%%%%%%%%%%%%%%%%%
 \end{eqnarray}
 Following the literature, the conformal and disformal coefficients are chosen \cite{Carsten2016a} :
\begin{eqnarray}  \label{E34} 
C(\phi) = \exp(2\alpha \kappa \phi)  \;, \hspace{1.0cm}
D(\phi) = \dfrac{\exp(2(\alpha + \beta) \kappa \phi)}{M^4}\; ,   \\  \label{E35}
\lambda_C \equiv - \dfrac{1}{\kappa}\dfrac{C_{,\phi}}{C} =-2\alpha \;, \hspace{0.5cm}
\lambda_D \equiv - \dfrac{1}{\kappa}\dfrac{D_{,\phi}}{D} =-2(\alpha + \beta)  \;, 
\end{eqnarray} 
where the parameters $\alpha$ and $\beta$ are constants. $M$ is the mass scale in the disformal coupling. In a purely conformal case, functions $D(\phi)$ and $ D_{,\phi}$ vanish.
%%%%%%%%%%%%%%%%%%%%%%%%%%%%%%%%%%%%%%%%%%%%%%%%%%%%%%%%
\newpage
\section{The dynamical systems approach}
%%%%%%%%%%%%%%%%%%%%%%%%%%%%%%%%%%%%%%%%%%%%%%%%%%%%%%%%
 The dynamical system of any cosmological model can be constructed by considering that certain properties (e.g. energy density) of the different fluids evolve with time $t$ or e-folds  $N = \ln a $.  In our case, we shall define the dimensionless dynamical variables \cite{Carsten2016a} as follows:
\begin{eqnarray}  \label{E36}
x^2 \equiv  \dfrac{\kappa^2\dot{\phi}^{2} }{3H^2} , \hspace{1cm}
y^2 \equiv  \dfrac{\kappa^2\rho_{\Lambda}}{3H^2} ,\hspace{1cm}
z^2 \equiv \dfrac{\kappa^2\rho_{c}}{3H^2}  ,\hspace{1cm}
\sigma \equiv  \dfrac{D(\phi)}{\kappa^2 C(\phi)}H^2 \;.
\end{eqnarray}
It follows that the set of equations, governing the evolution of these dynamical variables, can be inferred from the continuity equations \eqref{E14} and \eqref{E15}. At this point, let us emphasise that the dynamical variable $\sigma$ measures the strength of the disformal coupling. These variables are constrained to obey the Friedmann and Raychaudhuri equations \eqref{E27} yielding:
 \begin{eqnarray}
1+ w_{u}= x^2 + (1+ w_{u})y^2 + (1+w_{c})z^2\;, \hspace{0.1cm} \\
%\text{and} \hspace{0.1cm} 
\dfrac{ \;H^{\prime}}{H} =-\dfrac{3}{2}\left[x^2 + (1+w_{c})z^2\right] = -\dfrac{3}{2}(1+w_{\text{eff}})  \;, \label{E37}
\end{eqnarray}
 where $^{\prime}$ means derivative with respect to e-fold $N$. The $w_{\text{eff}}$ has been introduced as an effective equation of state, which must satisfy  $ w_{\text{eff}} <-1/3$ for accelerated expansion and $w_{\text{eff}} >-1/3$ for decelerated expansion. The set of first order differential equations for the dynamical system is obtained by taking the derivative of the defined variables \eqref{E36} and plugging into the continuity equations \eqref{E14} and \eqref{E15}:
 
\begin{eqnarray}\label{E38}
x^{\prime} &=& - \dfrac{ \;H^{\prime}}{H}x - \dfrac{3x}{2}\dfrac{(1+ w_{u})}{(1-3\tilde{w}z^2\sigma)} +  \dfrac{3\sqrt{3}\tilde{w}x^2z^2\sigma(\lambda_{C}-\lambda_{D})}{2(1-3\tilde{w}z^2\sigma)} + \dfrac{3z^2}{2x}\dfrac{(w_{c}-w_{u})(1+w_{c})}{(1-3\tilde{w}z^2\sigma)}  \;   
\nonumber\\
&& + \; \dfrac{(1+w_{c})}{(1-3\tilde{w}z^2\sigma)} \frac{\kappa}{2\sqrt{3}}\dfrac{Q}{H^2} \;,\\ [5pt] \label{E39}
 z^{\prime} &=& -\dfrac{ \;H^{\prime}}{H}z -\dfrac{3}{2}z(1+w_c) - \frac{\kappa}{2\sqrt{3}}\dfrac{Q}{H^2}\dfrac{x}{z} \;,  \\ \label{E40}
y^{\prime} &=& -\dfrac{ \;H^{\prime}}{H}y\;,  \\ \label{E41}
\sigma^{\prime} &=& \left((\lambda_C - \lambda_D)\sqrt{3}x  + 2\dfrac{ \;H^{\prime}}{H} \right)\sigma  \;.
 \end{eqnarray}
 %\frac{\kappa}{2\sqrt{3}}\dfrac{Q}{H^2}
%%%%%%%%%%%%%%%%%%%%%%%%%%%%%%%%%%%%%%%%%%%%%%%%%%%%
 After substituting the above dynamical variables \eqref{E36} into the interaction term \eqref{E33}, we obtain \eqref{E42}. This is a closed system because $\left(\frac{H^{\prime}}{H}\right)$ can be eliminated by using \eqref{E37}.
 \begin{eqnarray} \label{E42}
\hat{Q} \equiv \dfrac{\kappa Q}{H^2}&=& \dfrac{6z^2(1-3\tilde{w}\sigma z^2)\left[  \dfrac{\lambda_{C}}{2}(1-3w_c) + \dfrac{3}{2}\sigma x^2 (\lambda_{D} -2\lambda_{C}) + 3\sqrt{3}w_{c}\sigma x \right]}{(1-3\sigma x^2)(2-6\tilde{w}\sigma z^2) +3\sigma z^2(1+w_c)} +  \nonumber \\
&& \;  \dfrac{9\sqrt{3}\sigma z^2\left[ x(1+w_u)  + \sqrt{3}\tilde{w}\sigma x^2 z^2(\lambda_{D} -\lambda_{C}) - (w_{c}-w_{u})(1+ w_{c})\dfrac{z^2}{x} \right]}{(1-3\sigma x^2)(2-6\tilde{w}\sigma z^2) +3\sigma z^2(1+w_c)}\;.
 \end{eqnarray}
 These differential equations \eqref{E38}-\eqref{E42} govern the overall behaviour for a FLRW Universe. The interaction term $Q(\phi)$ is a function of $\lambda_C =-2\alpha $ and $ \lambda_D =-2(\alpha + \beta)$. Thus the choice of the functions $C(\phi)$ and $D(\phi)$ determines how the coupled fluid and the quintessence are interacting. Using the Friedmann constraint i.e. the first equation in \eqref{E37}, one can express $z$ in terms of $x,\;y$  and then substitute it in the evolution equations  \eqref{E38}-\eqref{E42} to reduce the dimensions of the dynamical system, with only $x,\;y$ and $\sigma$. This set of equations is more useful and straightforward to be solved, and we can always recover $z$ through the constraint equation whenever required. Finally, one can write the density parameter $\Omega_{\phi}$ from equation  \eqref{E26} and using  \eqref{E22} together with \eqref{E23} we can find the EOS for the quintessence $w_{\phi}$, in terms of the dynamical variables:
 \begin{eqnarray}  \label{E43}
\Omega_{\phi} \equiv   \dfrac{\kappa^2\rho_{\phi}}{3H^2} = \dfrac{x^2}{1+w_u} + y^2 + \dfrac{w_c -w_u}{1+w_u}z^2 = 1-z^2 \;\hspace{1cm}  \text{and} \;\hspace{1cm}  w_{\phi}= \dfrac{x^2 + y^2 -1}{1-z^2}\;.
 \end{eqnarray}
 %\dfrac{P_{\phi}}{\rho_{\phi}} &=& \dfrac{-y^2 -\dfrac{w_u x^2}{(1+w_u)} -z^2\left(\dfrac{w_u -w_c}{1+w_u}\right)}{(1-z^2)} 
%%%%%%%%%%%%%%%%%%%%%%%%%%%%%%%%%%%%%%%%%%%%%%%%%%%%
The solutions of this dynamical system with the differential equations \eqref{E38}-\eqref{E42} can be represented on a phase space of $x$, $z$, and $\sigma$, where each point denotes a specific state of the system (see Figures 2, 5 and 8).  However this phase space cannot be infinite as physical conditions must be imposed, for instance, the energy density can only be positive and therefore the existence of fixed points is limited by $ 0 <\Omega_{\phi} < 1 $. Also, the values of the disformal strength, $0  <\sigma <\infty $, is compactified by defining $\Sigma \equiv \arctan \sigma$. The resulting phase space, therefore, results in a compact set, defined by
\begin{eqnarray} \label{E44}
 -1 < x < 1 \hspace{0.6cm} \text{and} \hspace{0.6cm} z>0 \hspace{0.6cm}\text{and}\hspace{0.5cm} 0 < \Sigma < \pi/2 \;, \\ 
\hspace{0.2cm}\text{and} \hspace{0.2cm} 0 < x^2 + (1+w_u)y^2 + (w_c -w_u)z^2 < (1+w_u)  \;.  \nonumber
 \end{eqnarray}
We verify whether the transformation \eqref{E2}  is indeed real and invertible, by computing the Jacobian of the equation \eqref{E2}, which is expressed in terms of the dynamical variables below:
\begin{equation} \label{E45}
J = \dfrac{\sqrt{-g}}{\sqrt{-\tilde{g}}} =C^{2}\sqrt{1+\dfrac{D}{C}g^{ab}\D_{a}\phi\D_{b}\phi } = C^{2}\sqrt{1-3\sigma x^2}\;.
\end{equation}
It can be noted that if $3\sigma x^2 > 1 $,  the Jacobian $J$ is not real and in the limit $\sqrt{1-3\sigma x^2} \to 0$, a singularity appears in this phase space.  Therefore, the following condition must be imposed on the phase space:
\begin{equation} \label{E46}
3\sigma x^2 \leq 1  \;.
\end{equation}
The dynamical system \eqref{E38}-\eqref{E42} is also invariant under the simultaneous transformation of the parameters $(\alpha, \beta) \rightarrow (-\alpha,-\beta)$. Thus, the phase space is still fully described by taking only positive values of $(\alpha, \beta)$. 

%%%%%%%%%%%%%%%%%%%%%%%%%%%%%%%%%%%%%%%%%%%%%%%%%%%%
\section{Scenarios, Fixed points, and Trajectories}
%%%%%%%%%%%%%%%%%%%%%%%%%%%%%%%%%%%%%%%%%%%%%%%%%%%%
The dynamical system is described by the evolution equations \eqref{E38},\eqref{E39}, \eqref{E41}, the Friedmann constraint equation \eqref{E37}, two free parameters ($\alpha\;\text{and}\;\beta$) and the two fluids ($\rho_{\phi} \;\text{and}\;  \rho_c$) mimicking the cosmological constant $\Lambda$ and an uncoupled fluid $\rho_u$. The choice of EOS parameter $\tilde{w}$ and $\tilde{w}_u$  for the coupled fluid and uncoupled fluid respectively give rise to different scenarios: 
\begin{enumerate}
\item Scenario I : $\tilde{w} =0$ and $\tilde{w}_u =0$, implies that a pressureless fluid (e.g DM) is coupled to quintessence and both fluids are mimicking a $\Lambda \rm{CDM}$ background, where DM starts to dominate after matter-dark energy equality.
\item Scenario II: $\tilde{w} =1/3$ and $\tilde{w}_u =0$, implies that a relativistic fluid is coupled to quintessence and both fluids are mimicking a $\Lambda \rm{CDM}$ background.
\item Scenario III: $\tilde{w} =1/3$ and $\tilde{w}_u =1/3$, implies that  a relativistic fluid coupled to quintessence and both fluids are mimicking a background with a cosmological constant $\Lambda$ and an uncoupled relativistic fluid. An example of this scenario refers to relativistic neutrinos which are coupled to quintessence in the early universe as compared to uncoupled radiation.
\end{enumerate}

Our analysis of the dynamical system in each specific scenario i.e., after specifying the EOS parameters $\tilde{w}$ and $\tilde{w}_u$ is done by investigating two aspects:
\begin{enumerate}
\item The nature of the fixed points  -- which are static solutions  of the dynamical system, that correspond to $x^{\prime} = y^{\prime} = z^{\prime} = \sigma^{\prime} = 0$ \cite{Bahamonde2017}.
\item The trajectories on the phase space -- which are non-static solutions  of the dynamical variables. Those trajectories correspond to the evolution of the system from one cosmological state to another on the compactified phase space.
\end{enumerate}

According to linear stability theorem \cite{Bahamonde2017}, one can then determine how the system behaves around any point, by applying a Taylor expansion. The Jacobian matrix of the set of  three evolution equations is first worked out and then its corresponding three eigenvalues of that Jacobian matrix are evaluated for each of the fixed points. The nature of the fixed point is then determined as:
\begin{enumerate}
\item  If all three eigenvalues $ \Re(E_1) < 0 \; \text{and}\; \Re(E_2) < 0\; \text{and}\; \Re(E_3) < 0$, then the fixed point is stable,
\item  If all three eigenvalues $\Re(E_1) > 0 \; \text{and}\; \Re(E_2) > 0\; \text{and}\; \Re(E_3) > 0$, then the fixed point is unstable,
\item If any combination of three eigenvalues being positive and negative, then the fixed point is saddle,
\end{enumerate}
where $\Re(E_n)$ denote the real part of the $n^{th}$ eigenvalue. These eigenvalues can still have a dependence on the parameters $\alpha$ and $\beta$ and the conditions for which the fixed points are either stable, unstable or saddle, must be worked out. In other cases, where the eigenvalues are found to be null, the linear stability theorem is inadequate to determine their nature. Then the nature of the point is then evaluated numerically using the Lyapunov theorem  \cite{Bahamonde2017} to investigate the asymptotic and global stability.

Given certain initial conditions and choice of parameters ($\alpha$ and $\beta$), the nature of the trajectories of this phase space can be determined. In what follows we determine the evolution of the dynamical variables $x, z$ and $\sigma$ for the two following cases:
\begin{enumerate}
\item Fixed parameters ($\alpha$ and $\beta$) but different Initial Conditions (IC).
\item Fixed Initial Conditions (IC) but different parameters ($\alpha$ and $\beta$).
\end{enumerate}
Our approach is to investigate all the different theoretical possibilities for (a) the trajectories,  (b) the effect of disformal coupling constant on the evolution of the Universe, and (c) the sensitivity of the dynamical system to the initial conditions. In the upcoming sections VI-VIII, we will analyse the nature of the fixed points and trajectories for three different scenarios in both conformal and disformal framework.

%\newpage
\section{Conformal Framework}
We investigate the fixed points and the phase portrait of the dynamical system in a conformal framework (i.e. $C(\phi) \neq 1$ in equation \eqref{E1}) for the three studied scenarios. In a purely conformal framework, the disformal coefficient and its derivative vanish $D(\phi)=D_{,\phi}=0$ and the variable $\sigma$ also vanishes. We obtain the differential equations for $x$ and $z$ as below for the three studied scenarios. 

\begin{enumerate}
\item In Scenario I ($\tilde{w} =0$ and $\tilde{w}_u =0$), the interaction between quintessence and is still present but simplies reduces to $Q =  \alpha \rho\nabla^{b}\phi$ \cite{Amendola1999}, where the parameter $\alpha$ is the conformal coupling constant. 
\begin{eqnarray}  \label{E47}
 x^{\prime} &=& \dfrac{3x}{2}(x^2 + z^2)-\dfrac{3x}{2} - \dfrac{\kappa}{2}\sqrt{3}\alpha z^{2} \;,  
\\  \label{E48}
 z^{\prime} &=& \dfrac{3z}{2}(x^2 + z^2) - \dfrac{3z}{2} + \dfrac{\kappa}{2}\sqrt{3}\alpha xz \;.
 \end{eqnarray}
 \item When $\tilde{w} =1/3$, the quintessence and the radiation do not interact at all with each other, and so the interaction term $Q(\phi)$, vanishes as well for both scenarios II and III. In Scenario II: ($\tilde{w} =1/3$ and $\tilde{w}_u =0$), the differential equations becomes:
\begin{eqnarray}  \label{E57S3}
 x^{\prime} &=&\dfrac{3}{2} x \left(x^2+\dfrac{4 z^2}{3}\right)-\frac{3 x}{2} +\frac{2 z^2}{3 x}\;, 
 \\  \label{E58S3}
 z^{\prime} &=& \dfrac{3}{2} z \left(x^2+\dfrac{4 z^2}{3}\right)-2 z\;.
 \end{eqnarray}
 
 \item In Scenario III ($\tilde{w} =1/3$ and $\tilde{w}_u =1/3$), the differential equations becomes:
\begin{eqnarray}  \label{E62b}
 x^{\prime} &=&\frac{3}{2} x \left(x^2+\frac{4 z^2}{3}\right)-2 x\;,
 \\  \label{E63}
 z^{\prime} &=&\frac{3}{2} z \left(x^2+\frac{4 z^2}{3}\right)-2 z\;.
 \end{eqnarray} 
\end{enumerate}

\begin{figure}[h!] 
\centering 
\includegraphics[scale=0.4]{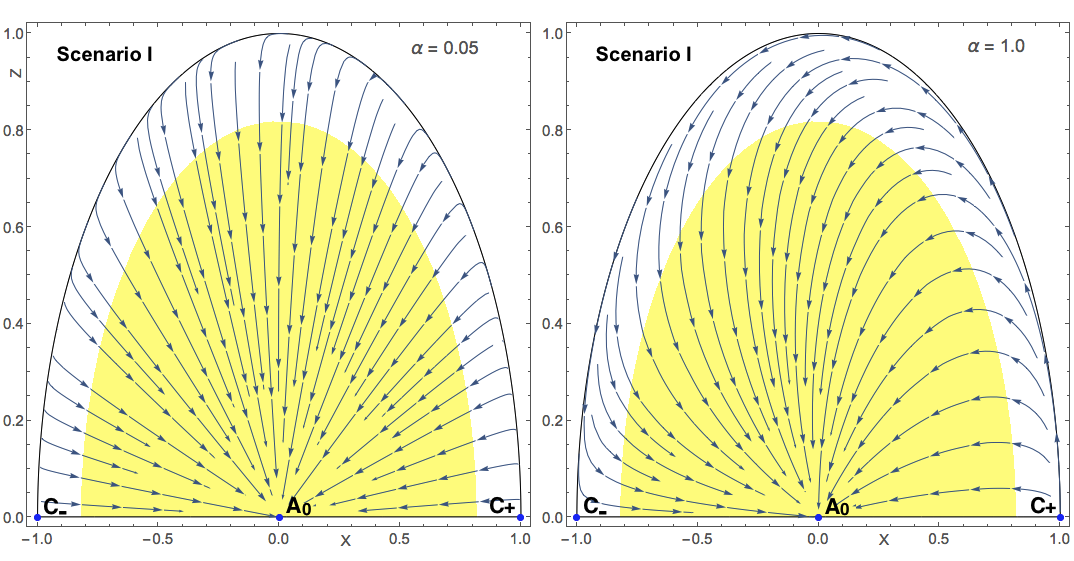} 
\includegraphics[scale=0.375]{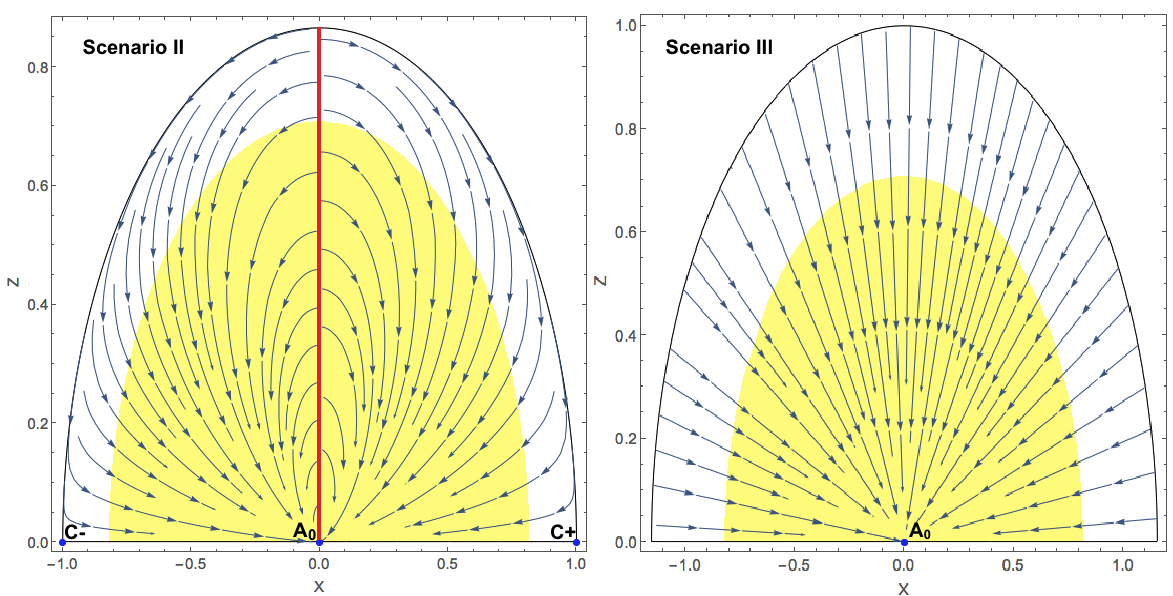} 
\caption{\label{fig:stream-sc1}Trajectories of the dynamical system  \eqref{E47}-\eqref{E48} in the scenario I (i.e. $\tilde{w} =0$ and $\tilde{w}_u =0$), are shown in the top panels, with  $\alpha = 0.05$ in the top left and $\alpha = 1.0$ in the top right respectively. In the top panels, the Friedmann constraint \eqref{E37} becomes $1 = x^2 + z^2$ and represents the half-circle. Trajectories of the dynamical system  \eqref{E57S3}-\eqref{E58S3} in the conformal scenario II (i.e $\tilde{w} =1/3$ and $\tilde{w}_u =0$), are shown in the bottom left panel, where the red line is equilibrium line consisting of repeller fixed points and it separates $x<0$ and $x>0$. In the bottom left panel, the Friedmann contraint \eqref{E37} becomes $3 = 3x^2 + 4z^2$ and represents the half-ellipse. The trajectories of the dynamical system \eqref{E62b}-\eqref{E63}  in the conformal scenario III (i.e $\tilde{w} =1/3$ and $\tilde{w}_u =1/3$), are shown in the bottom right panel. In conformal scenario III, the Friedmann contraint \eqref{E37} becomes $4 = 3x^2 + 4z^2$ and represents the half-ellipse. In all panels, the blue dots represents the fixed points i.e. $A_{0}$ and $(C_{\pm})$, which are the attractor and saddle points respectively. The yellow shaded region corresponds to a region of accelerated expansion, satisfying  $ w_{\text{eff}} <-1/3$. }
\end{figure}

 The phase portraits of the  dynamical system \eqref{E47}-\eqref{E48}, as well as \eqref{E57S3}-\eqref{E58S3} and  \eqref{E62b}-\eqref{E63} was computed, as shown in Figure \ref{fig:stream-sc1}. The following points were noted:
\vspace{-0.2cm}
\begin{enumerate}
\item Point $A_0$, i.e. ($0,0$), is a scalar field dominated fixed point which exists in scenario I, II, and III. It is located at the origin of the phase space. It is an attractor when evaluated according to Lyapunov theorem. Most of the trajectories seem to end at this fixed point. It always exists since it has no dependence on parameters $\alpha$. It satisfies the acceleration condition of $ w_{\text{eff}} <-1/3$. 

\item Points $(C_{\pm})$, i.e. ($\pm1,0$), which are two kinetic-dominated fixed points, which exist in scenario I, II, but not III. They are located on the $xz$  plane at unit length of $x$. They have no dependence on the parameter $\alpha$ and always exists. In scenario I, when $\alpha >0 $, Point  $(C_{+})_{\textbf{I}}$ is unstable and Point $(C_{-})_{\textbf{I}}$ is saddle and when $\alpha < 0$, Point $(C_{+})_{\textbf{I}}$ is unstable and Point $(C_{-})_{\textbf{I}}$ is saddle. In scenario II, Points $(C_{\pm})_{\textbf{I}}$ are both saddle.

\item In scenario II, the line $x = 0$ is an equilibrium line, which consist of repeller fixed points and it separates $x<0$ and $x>0$.
\end{enumerate}

In scenario I, the values of $\alpha$ were constrained to be $\alpha = \pm 0.079^{+0.016}_{-0.019}$ for an exponential potential, as mentioned in \citep{Barros2018}. Thus, two phase portraits were produced with $\alpha = 0.05$ and $\alpha = 1.0$ for illustrative purposes. In scenario II and III, the dynamical systems \eqref{E57S3}-\eqref{E58S3} and  \eqref{E62b}-\eqref{E63} does not depend on the parameters $\alpha$ and $\beta$ at all.

%%%%%%%%%%%%%%%%%%%%%%%%%%%%%
\section{Disformal Framework}
%%%%%%%%%%%%%%%%%%%%%%%%%%%%%
We extend now our analysis of the fixed points and the phase portrait to a disformal framework for the three studied scenarios. The dynamical variable $\sigma$ and the disformal constant $\beta$ now play an important role. The form of the differential equations for \eqref{E38}, \eqref{E39},  \eqref{E41} and the reduced interaction term \eqref{E42} can be written for the respective scenarios as follows:
%&&&&&&&&&&&&&&&&&&&&&&&&&&&&&&&&&&&&&&&&&&&&
\begin{enumerate}
\item In scenario I ($\tilde{w} =0$ and $\tilde{w}_u =0$): The dynamical system, described by \eqref{E49}, \eqref{E50} and \eqref{E51},  has 9 fixed points, which are summarized in Table \ref{table:points-sc1} with respective eigenvalues in Table \ref{table:eigen-sc1}.
\begin{eqnarray}  \label{E49}
 x^{\prime} &=& \dfrac{3x}{2}(x^2 + z^2)-\dfrac{3x}{2} +\dfrac{ \hat{Q}}{2\sqrt{3}}\;,  
\\  \label{E50}
 z^{\prime} &=& \dfrac{3z}{2}(x^2 + z^2) - \dfrac{3z}{2} -\dfrac{\hat{Q}}{2\sqrt{3}}\left(\dfrac{x}{z}\right)\;, 
 \\ \label{E51}
 \sigma^{\prime} &=& \left[\sqrt{3} x \left(\lambda _C-\lambda _D\right)-3 \left(x^2+z^2\right)\right] \sigma\;,
 \\ \label{EQ1}
 \hat{Q } &=& \frac{3 z^2 \left[\lambda _C \left(1-6 \sigma  x^2\right)+3 \sigma  x \left(x \lambda _D+\sqrt{3}\right)\right]}{-6 \sigma  x^2+3 \sigma  z^2+2}\;.
 \end{eqnarray}
 
%&&&&&&&&&&&&&&&&&&&&&&&&&&&&&&&&&&&&&&&&&&&&
\item In scenario II ($\tilde{w} =1/3$ and $\tilde{w}_u =0$): The dynamical system, described by \eqref{E57}, \eqref{E58} and \eqref{E59}, has 5 fixed points which are in Table \ref{table:points-sc2} with their Eigenvalues in Table \ref{table:eigen-sc2}.
\begin{eqnarray}  \label{E57}
 x^{\prime} &=&+\frac{3}{2} x \left[x^2 +\frac{z^2}{3} \left(4-3 \sigma  x^2\right) \right] - \frac{3 x}{2 \left(1-\sigma  z^2\right)} + \frac{z^2 \left(4 - 3 \sigma  x^2\right) \left(1-3 \sigma  x^2\right)}{6 x \left(1 -\sigma  z^2\right)} \hspace{1.0cm} \\
&&\; +  \frac{\sqrt{3} \sigma  x^2 z^2 \left(\lambda _C-\lambda _D\right)}{2 \left(1-\sigma  z^2\right)}  +  \dfrac{\left(4-3 \sigma  x^2\right)}{6 \sqrt{3} \left(1-\sigma  z^2\right)}\hat{Q}\;,  \nonumber \\  \label{E58}
 %%%%%%%%%%%%%%%%%%%%%%%%%%%%%
 z^{\prime} &=& +\frac{3}{2} z \left[x^2 +\frac{z^2}{3} \left(4-3 \sigma  x^2\right) \right] - \frac{z}{2}\left(4-3 \sigma  x^2\right) -\frac{x}{2 \sqrt{3} z}\hat{Q}  \;, \\\label{E59}
%%%%%%%%%%%%%%%%%%%%%%%%%%%%%
 \sigma^{\prime} &=& \sigma  \left\lbrace\sqrt{3} x \left(\lambda _C-\lambda _D\right)- \left[3x^2 +z^2\left(4-3 \sigma  x^2\right)\right]\right\rbrace\;, \\ \label{E60}
%%%%%%%%%%%%%%%%%%%%%%%%%%%%%
 \hspace{0.5cm} \hat{Q}&=& \frac{\sigma  z^2 \left\lbrace 9 x^3 (\lambda _D-\lambda _C)+\sqrt{3} \left[9 \sigma  x^4 \left(\sigma  z^2-2\right)+3 x^2 \left(3 \sigma  z^2+5\right)-4 z^2\right]\right\rbrace}{\sigma  z^2 \left(4-3 \sigma  x^2\right)+2 \left(1-3 \sigma  x^2\right) \left(1-\sigma  z^2\right)}\;.
 %%%%%%%%%%%%%%%%%%%%%%%%%%%%%
 \end{eqnarray}
 \item In scenario III ($\tilde{w} =1/3$ and $\tilde{w}_u =1/3$), The dynamical system \eqref{E64}-\eqref{E67},  has 7 fixed points, which are summarized in Table \ref{table:points-sc3}, together with their eigenvalues.
 \begin{eqnarray}  \label{E64}
 x^{\prime} &=&+\frac{3}{2} x \left[x^2 +\frac{z^2}{3} \left(4-3 \sigma  x^2\right) \right] - \frac{2 x}{\left(1-\sigma  z^2\right)} -\frac{\sigma xz^2 \left(4-3 \sigma  x^2\right)}{2 \left(1 -\sigma  z^2\right)} + \nonumber \\&&\;  \frac{\sqrt{3} \sigma  x^2 z^2 \left(\lambda _C-\lambda _D\right)}{2 \left(1-\sigma  z^2\right)}  +  \dfrac{\left(4-3 \sigma  x^2\right)}{6 \sqrt{3} \left(1-\sigma  z^2\right)}\hat{Q}\;,  \\
\label{E65}
%%%%%%%%%%%%%%%%%%%%%%%%%%%%%
 z^{\prime} &=& +\frac{3}{2} z \left[x^2 +\frac{z^2}{3} \left(4-3 \sigma  x^2\right) \right] - \frac{z}{2}\left(4-3 \sigma  x^2\right) -\frac{x}{2 \sqrt{3} z}\hat{Q}  \;, 
%%%%%%%%%%%%%%%%%%%%%%%%%%%%%
 \end{eqnarray} 
 %%%%%%%%%%%%%%%%%%%%%%%%%%
\begin{eqnarray}\label{E66}
 \sigma^{\prime} &=& \sigma  \left\lbrace\sqrt{3} x \left( \lambda _C-\lambda _D\right)- \left(3x^2 +z^2\left[4-3 \sigma  x^2\right)\right]\right\rbrace \;, \\ \label{E67}
%%%%%%%%%%%%%%%%%%%%%%%%%%%%%
 \hspace{0.5cm} \hat{Q}&=& \frac{3 \sigma  x z^2 \left\lbrace 3 x( \lambda _D -\lambda _C)+\sqrt{3} \left[3 \sigma  x^2 \left(\sigma  z^2-2\right)+2 \sigma  z^2+6\right]\right\rbrace}{3 \sigma  x^2 \left(\sigma  z^2-2\right)+2 \sigma  z^2+2}\;.
 %%%%%%%%%%%%%%%%%%%%%%%%%%%%%
\end{eqnarray}
\end{enumerate}

All the disformal fixed points and features of the general dynamical system \eqref{E38}-\eqref{E42} are iterated below and the following remarks are made:
\begin{enumerate}
\item Point $A_0$, is a scalar field dominated fixed point, which exists in Scenario I, II and III (same as in the conformal framework). It is located at the origin of the phase space. It is an attractor when evaluated according to Lyapunov theorem. Most of the trajectories end at this fixed point. It always exists since it does not depend on parameters $\alpha$ and $\beta$. It satisfies the acceleration condition of $ w_{\text{eff}} <-1/3$. The fixed point $A_0$ has a null eigenvalue and it was found to be stable by using Lyapunov theorem instead. 

\item Point $(C_{\pm})$, i.e. ($\pm1,0,0$) are two kinetic-dominated fixed points which exist in scenario I and II, but not scenario III. They are located on the $xz$  plane at unit length of $x$. They are similar as in the conformal framework except that $\beta$ can now influence its nature. 

%----------------------------------------------------------------
\begin{enumerate}[label=(\roman*)]
\item In scenario I, they always exist for all values of $\alpha$ and $\beta$ but their different choices determine whether the fixed points are either unstable or saddle points. For Point $(C_{+})_{\textbf{I}}$, the  conditions are the following:
\begin{small}
\begin{eqnarray} \label{E52}
\text{Unstable:}&\Rightarrow & \hspace{0.2cm} \beta >\frac{\sqrt{3}}{2} \hspace{0.2cm} \text{and} \hspace{0.2cm}  \alpha > 0\;.\\
\text{Saddle:}   &\Rightarrow &\hspace{0.2cm} \beta <\frac{\sqrt{3}}{2} \hspace{0.2cm} \text{and} \hspace{0.2cm} \alpha <0 \hspace{0.2cm} \text{OR} \hspace{0.2cm} \beta <\frac{\sqrt{3}}{2} \hspace{0.2cm} \text{and} \hspace{0.2cm}  \alpha >0 \hspace{0.2cm} \text{OR}\hspace{0.2cm} \beta >\frac{\sqrt{3}}{2} \hspace{0.2cm} \text{and} \hspace{0.2cm} \alpha <0\;.  \nonumber
\end{eqnarray}
\end{small}

For Point $(C_{-})_{\textbf{I}}$, the  conditions are the following:
 \begin{small}
\begin{eqnarray} \label{54}
\text{Unstable:}&\Rightarrow & \hspace{0.2cm} \beta <-\frac{\sqrt{3}}{2} \hspace{0.2cm} \text{and} \hspace{0.2cm}  \alpha <0\;.\\ 
\text{Saddle:}   &\Rightarrow &\hspace{0.2cm} \beta >-\frac{\sqrt{3}}{2} \hspace{0.2cm} \text{and} \hspace{0.2cm} \alpha >0 \hspace{0.2cm} \text{OR} \hspace{0.2cm} \beta >-\frac{\sqrt{3}}{2}\hspace{0.2cm} \text{and} \hspace{0.2cm}  \alpha <0 \hspace{0.2cm} \text{OR}\hspace{0.2cm} \beta <-\frac{\sqrt{3}}{2} \hspace{0.2cm} \text{and} \hspace{0.2cm} \alpha >0\;.  \nonumber
 \end{eqnarray}
 \end{small}

\item In scenario II, the fixed points always exist as saddle points and the conditions are:
\begin{small}
\begin{eqnarray} \label{E61}
\text{For $(C_{+})_{\textbf{II}}$ } &:& \beta <\frac{\sqrt{3}}{2} \hspace{0.2cm} \text{OR}  \hspace{0.2cm} \beta >\frac{\sqrt{3}}{2}\;. \\ \label{E62}
\text{For $(C_{-})_{\textbf{II}}$ } &:& \beta >-\frac{\sqrt{3}}{2} \hspace{0.2cm} \text{OR}  \hspace{0.2cm} \beta <-\frac{\sqrt{3}}{2}\;.
\end{eqnarray}
\end{small}
\end{enumerate}
%----------------------------------------------------------------
\item Point $(B_{\pm})$ are two  disformal scaling points which are possible solutions in scenario I and II. They both can only exist if either $\beta \leq -\frac{\sqrt{3}}{2}$ or $\beta \geq \frac{\sqrt{3}}{2}$. The location of the two fixed points depends on the parameter $\beta$. When $\beta = 0$ in conformal case, and those fixed points do not exist at all, so there is no issue of singularity at all. But unfortunately, the Points $(B_{\pm})_{\textbf{I}}$ and $(B_{\pm})_{\textbf{II}}$  are unable to satisfy the condition \eqref{E46}, which means the points are neither real nor physical points. In scenario I, these fixed points could only be saddle points and the conditions are: 
\begin{eqnarray}\label{EPtB}
0<\beta <\frac{\sqrt{3}}{2}  \;\text{if}\; \alpha<0 \; \text{and} \;-\frac{\sqrt{3}}{2}<\beta <0 \;\text{if}\; \alpha>0\;\\ \text{OR} \;\beta >\frac{\sqrt{3}}{2}  \;\text{if}\; \alpha<0 \;\text{and} \;\beta <-\frac{\sqrt{3}}{2}  \;\text{if}\; \alpha>0.
\end{eqnarray}
%---------------
\item Points $(D_{\pm})_{\textbf{III}}$ i.e. ($0, \pm1,0$), are two radiation-dominated fixed points which exists only in scenario III. They are located on the $xz$  plane at $z=\pm1$. It is an unstable point when evaluated according to Lyapunov theorem. The point $(D_{-})_{\textbf{III}}$ is unphysical because the energy density must be positive. The point $(D_{+})_{\textbf{III}}$ always exists as an unstable point.  Since its effective EOS $w_{\text{eff}} = 1/3$, this fixed points leads to a decelerated expansion.
%---------------
\item Points $(S_{\pm})$, i.e. $(0, 0,\pm \infty)$, arise due to the restriction imposed by the condition \eqref{E46} since they lie at $x=0$ , and $\sigma =\pm \infty$. Hence these fixed points exist in each scenario I, II and III. Since the phase portrait is compactified and defined by \eqref{E44}, and by the definition of the $\sigma$ variable is only positive, the Point $(S_{-})$ is unphysical. Point $(S_{+})$ is a Saddle.
 %---------------
\item Points $(T_{\pm})$, i.e. $(0, +1 ,\pm \infty)$,  arise due to the restriction imposed by the condition \eqref{E46} since they lie at $x=0$ , and $\sigma =\pm \infty$. These fixed points exist in scenario I and  III only but not in scenario II. Since the phase portrait is compactified and defined by \eqref{E44}, and by the definition of the $\sigma$ variable is only positive, Point $(T_{-})$ is unphysical. Point $(T_{+})$ is a repeller.
%---------------
\item  The line $L_{\rm{E}}$ i.e. $(x=0,z=0, \forall\; \sigma)$ constitutes a line of equilibrium points and it connects the fixed points $A_0$ and $(S_{+})$. This line exists in each scenario I, II and III.
%---------------
\item In scenario II, the  $z\sigma$ surface at $(x=0\; \forall\; \sigma)$ forms a topological surface of repeller points. 
%---------------
\end{enumerate}
%---------------
\section{Disformal Trajectories}
\subsection*{8.1.1  Disformal Scenario I: Fixed ($\alpha,\beta$) values but various Initial Conditions}
\begin{figure}[h!] 
\includegraphics[scale=0.4]{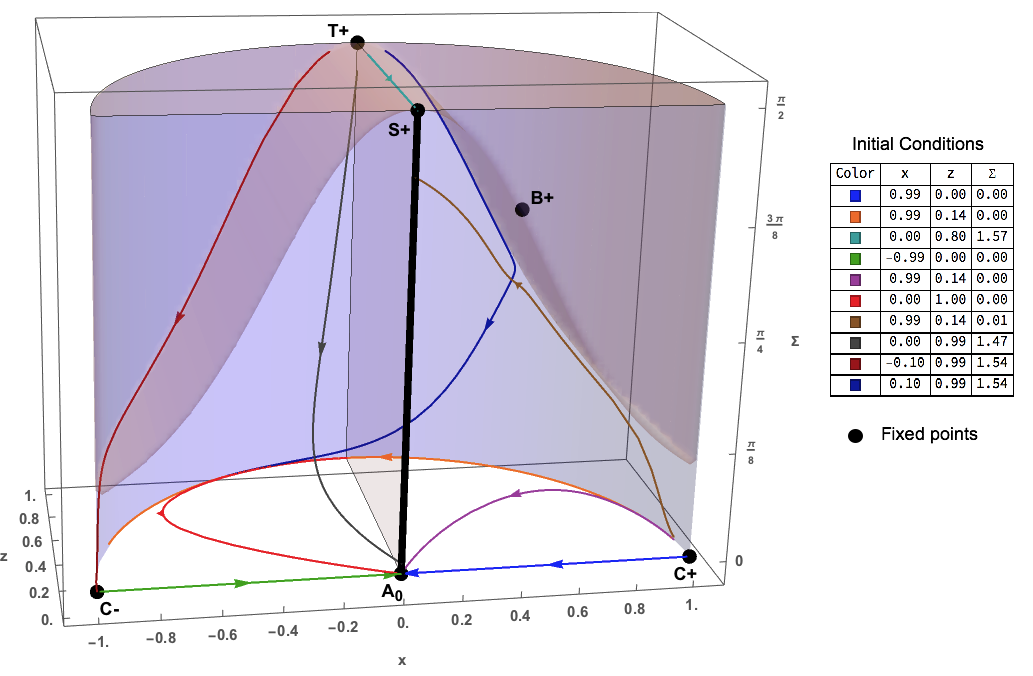} 
\caption{\label{fig:ic-sc1}This figure shows scenario I, where the quintessence is disformally coupled to a pressureless fluid (i.e $\tilde{w} =0$ and $\tilde{w}_u =0$) with $\alpha = 0.05, \;\text{and}\; \beta = 1.5$  and the $x\sigma z$ phase portrait is shown with the different trajectories, which correspond to different chosen initial conditions (including the fixed points $(C_{\pm})_{\textbf{I}}$ and $(T_{+})_{\textbf{I}}$).At the boundary of the phase space, the Friedmann contraint \eqref{E37} becomes $1 = x^2 + z^2$ and represents the surface for the half cylinder. The black solid line is the equilibrium line $L_{\rm{E}}$ i.e. $(x=0,z=0, \forall\; \sigma)$. The $xz$ plane ($\sigma =0$) is the conformal invariant submanifold and the fixed point $A_0$ is an attracting end point. The shaded region corresponds to the forbidden region, where condition \eqref{E46} is not satisfied.}
\end{figure}

%\vspace{0.1cm}
After having studied the fixed points, we move to the analysis of the non-static solutions in this scenario I (i.e $\tilde{w} =0$ and $\tilde{w}_u =0$). In order to fix ($\alpha,\beta$) values, we consider that $\beta \geq \sqrt{3}/2$ for Point $(B_{+})_{\textbf{I}}$ to exist and that $\alpha =0.05$ as constrained in \citep{Barros2018}. Let us now fix the values $\alpha = 0.05, \;\text{and}\; \beta = 1.5$ and analyse how our dynamical system evolves from different sets of initial conditions  (IC). Under the evolution of the dynamical system, the initial condition pre-determines the endpoints. The following was found from analysing figure  \ref{fig:ic-sc1}:
\begin{enumerate}
\item The dynamical system flows from one fixed point to another within this closed compactified space, where the Points $(S_{+})_{\textbf{I}}$ and $(T_{+})_{\textbf{I}}$ are asymptotes, and trajectories tend towards that limit.
\item If the starting point is near or on the $xz$  plane and satisfying $x^2 + z^2 \leq 1$, the end point is only the attractor point $A_{0}$, as in the conformal case. The conformal invariant sub-manifold is preserved.

\item The dynamical system is sensitive to the chosen initial conditions. i.e., different initial conditions around the proximity of the fixed points can lead to different endpoints. For example, when the IC is chosen near $(C_{+})_{\textbf{I}}$, the dynamical system can end up at either $A_{0}$ or on the equilibrium line near$(S_{+})_{\textbf{I}}$. Similarly, IC is chosen near $(T_{+})_{\textbf{I}}$, the endpoints  could be either at $A_{0}$ or $(S_{+})_{\textbf{I}}$ or on the equilibrium line $L_{\rm{E}}$. $(C_{-})_{\textbf{I}}$ is a saddle fixed point where the dynamical system can be momentarily at rest such that point $A_{0}$ is the final end point.
\end{enumerate}

\subsection*{8.1.2  Disformal Scenario I: Fixed Initial Conditions but various ($\alpha, \beta$) values}
\begin{figure}[h!]
\centering
\includegraphics[scale=0.39]{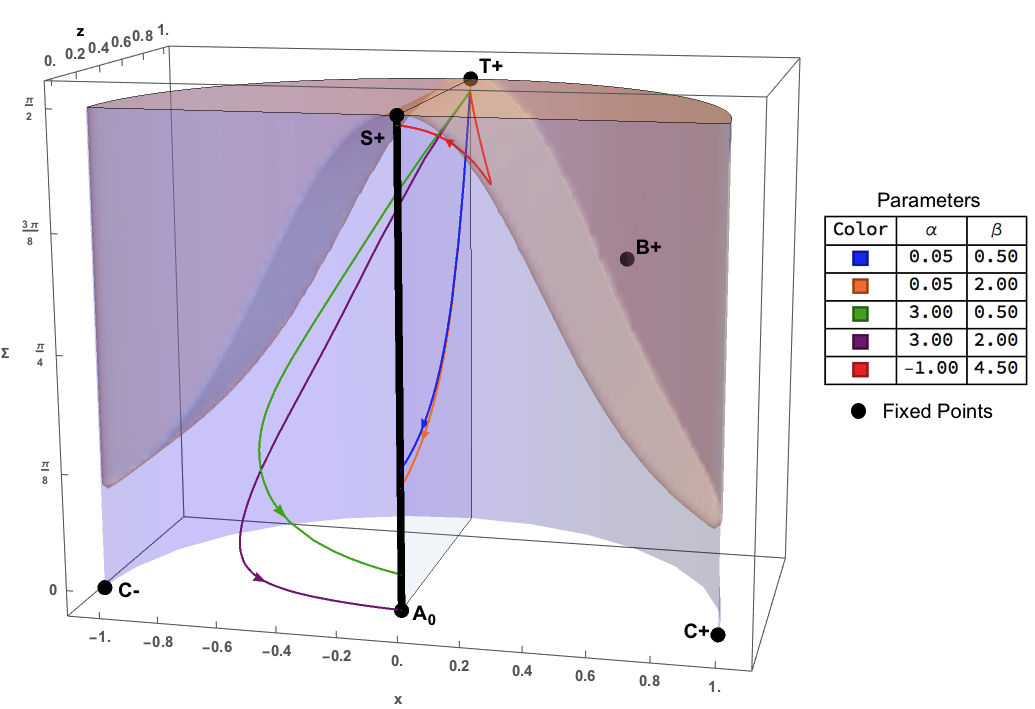} 
\caption{\label{fig:param-sc1}This figure shows scenario I, where the quintessence is disformally coupled to pressureless fluid (i.e. $\tilde{w} =0$ and $\tilde{w}_u =0$). Given $(T_{+})_{\textbf{I}}$ as the same initial condition, the evolution of the dynamical system  \eqref{E49}, \eqref{E50} and \eqref{E51} depend on the choice of parameters ($\alpha$ and $\beta$). At the boundary of the phase space, the Friedmann contraint \eqref{E37} becomes $1 = x^2 + z^2$ and represents the surface for the half cylinder. The different trajectories  in the $xz\sigma$ phase space correspond to different combination of parameters.}
\end{figure}

\vspace*{0.5cm}
At this stage, it makes sense to choose initial conditions (IC) close to a fixed point, because then, one can already know whether the chosen IC behaves like a saddle or a repeller. Moreover, it is preferable not to choose the IC near an attractor. The aim here is to investigate any deviation from the conformal case. Thus, it is in our interest to chose the IC set near to the repeller point $(T_{+})_{\textbf{I}}$. i.e ($x_{\rm{ic}}=0.0,z_{\rm{ic}}=+1.0$, and $\Sigma_{\rm{ic}} = \pi/2$). The different combination of  parameters ($\alpha,\beta$) can now be tested in order to determine how they affect the evolution of the dynamical system and are shown in figure \ref{fig:param-sc1}. 

We make the following remarks based on figure \ref{fig:param-sc1}:
\begin{enumerate}
\item For $\alpha >0$, the trajectories flow in the region bounded by $x<0$  and for $\alpha <0$,  the trajectories flow in the region bounded $x>0$. A larger value of $\alpha$ causes the trajectory to pass nearer to the conformal point$(C_{\pm})_{\textbf{I}}$ before ending at the attractor  $A_0$ or on the equilibrium line $L_{\rm{E}}$.
\item  For $\alpha <0$ and for $\beta >0$, a larger value of $\beta$ causes the trajectory to end nearer the disformal saddle point $(S_{+})_{\textbf{I}}$, otherwise for smaller value of $\beta$, it ends on equilibrium line $L_{\rm{E}}$, nearer to the attractor $A_0$.
\item  For $\alpha >0$ and for $\beta >0$, any value of $\beta$ causes the trajectory to repel away from the disformal saddle point $(S_{+})_{\textbf{I}}$ and nearer to the attractor $A_0$.
\item  For $\alpha <0$ and for $\beta <0$, any value of $\beta$ causes the trajectory to repel away from the disformal saddle point $(S_{+})_{\textbf{I}}$ and end unto the attractor $A_0$.
\item The parameters ($\alpha,\beta$) determine how much the dynamical system  \eqref{E49}, \eqref{E50} and \eqref{E51} gets influenced by either the conformal fixed points $(C_{\pm})_{\textbf{I}}$ or disformal fixed point $(S_{+})_{\textbf{I}}$, as it evolves on its trajectory within the phase space.
\end{enumerate}

\vspace*{0.2cm}
\subsection*{8.2.1 Disformal Scenario II: Fixed ($\alpha,\beta$) values but various Initial Conditions}
\vspace*{0.2cm}

 The next two subsections now consider the phase portrait in the scenario II (i.e $\tilde{w} =1/3$ and $\tilde{w}_u =0$). Let us now fix the values  $\alpha = 0.05,\; \beta = 1.5$  to same choice as previous scenarios for sake of comparison and analyse the trajectories from different sets of initial conditions.

\begin{figure}[h!] 
\centering
\includegraphics[scale=0.4]{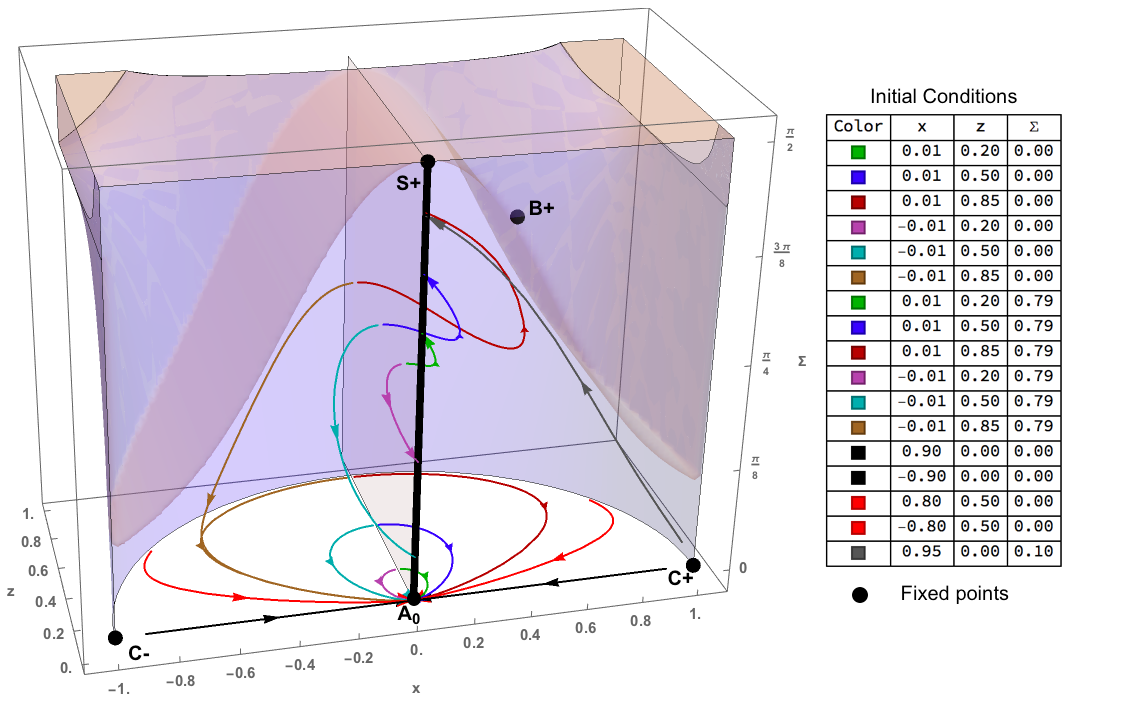} 
\caption{\label{fig:ic-sc2}This figure shows scenario II, where the quintessence is disformally coupled to relativistic fluid (i.e $\tilde{w} =1/3$ and $\tilde{w}_u =0$) with $\alpha = 0.05, \;\text{and}\; \beta = 1.5$  and the $x\sigma z$ phase portrait is shown with the different trajectories, which correspond to different chosen initial conditions from the repeller $z\sigma$ plane surface. The blue surface represents the boundary of the phase space and satisfies the Friedmann contraint \eqref{E37}, which becomes $3 x^2 \left(\sigma z^2-1\right)+3=4 z^2$. The black solid line is the equilibrium line $L_{\rm{E}}$ i.e. $(x=0,z=0, \forall\; \sigma)$. The $xz$ plane ($\sigma =0$) is the conformal invariant submanifold and the fixed point $A_0$ is an attracting end point. The $z\sigma$ plane constitute a topological separation between $x<0$ and $x>0$. The shaded region correspond to the forbidden region as where the condition \eqref{E46} is not satisfied.}
\end{figure} 
 
%\newpage
The following remarks were noted based on Figure \ref{fig:ic-sc2}:
\begin{enumerate}
\item We recover the conformal behaviour (compare with bottom left panel of Figure  \ref{fig:stream-sc1}) on the plane of $\sigma = 0$ as expected. The dynamical variable $\Sigma$ gives an additional degree of freedom for the dynamical system to evolve. The equilibrium line of repeller points in  Figure \ref{fig:ic-sc2} extends to a $z\sigma$ plane constitutes a topological separation between $x<0$ and $x>0$.

\item  All the trajectories starting on that $z\sigma$ plane  end on the equilibrium line $L_{\rm{E}}$, which connects both $(S_{+})_{\textbf{II}}$ and attractor point $A_0$.

\item If IC are set close to $z\sigma$ surface for $x>0$ and further away from the equilibrium line $L_{\rm{E}}$, the trajectory gets attracted to the saddle point $(S_{+})_{\textbf{II}}$ than $A_0$ before ending onto the equilibrium line $L_{\rm{E}}$.

\item If IC are set close to $z\sigma$ surface for $x<0$ and further away from the equilibrium line $L_{\rm{E}}$, the trajectory gets repelled by the saddle point $(S_{+})_{\textbf{II}}$ but also attracted to the attractor point $A_0$  before ending onto the equilibrium line $L_{\rm{E}}$.

\item For $z=0$, and $x<0\; \text{and}\;  \forall\; \Sigma$, all trajectories end up at the attractor  $A_0$. For $z=0$, and $x>0 \; \text{and}\;  \Sigma > 0$, all trajectories end up closer to the saddle point  $(S_{+})_{\textbf{II}}$, but if $\Sigma = 0$, the trajectory will end up at the attractor  $A_0$ as expected from the conformal behaviour.
\end{enumerate}

\subsection*{8.2.2 Disformal Scenario II: Fixed Initial Conditions but various ($\alpha, \beta$) values}
As done in subsection 8.1.2, in order to describe the disformal coupling, it makes sense to choose initial conditions (IC) close to a disformal fixed point.

\begin{figure}[h!]
\includegraphics[scale=0.375]{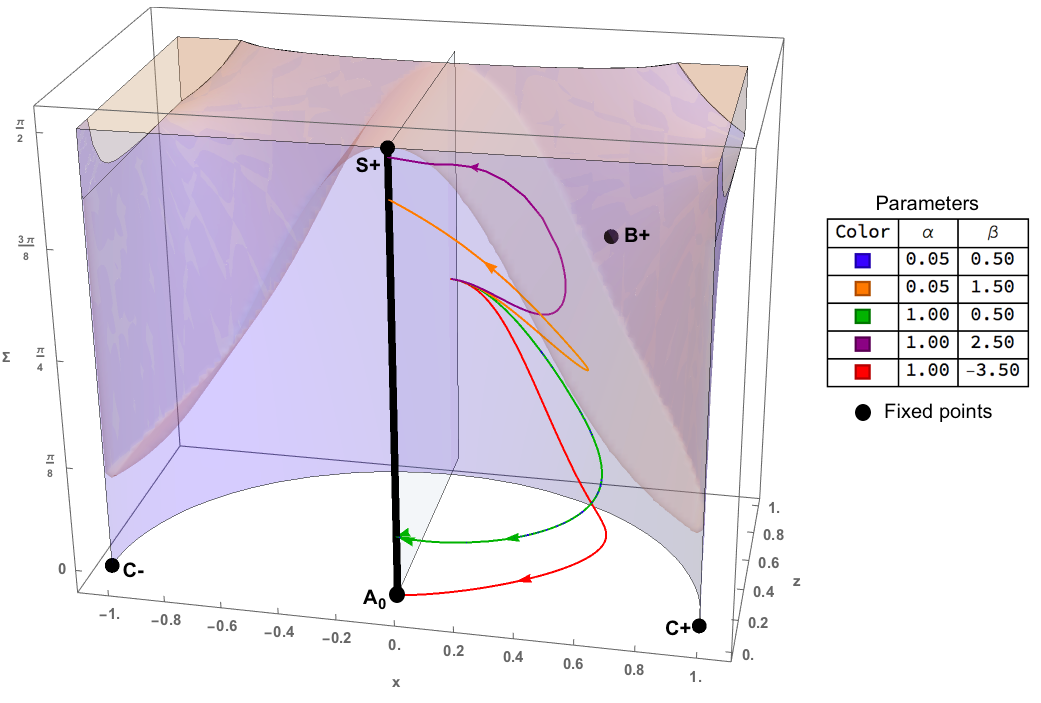} 
\caption{\label{fig:ic-sc5}This figure shows scenario II, where the quintessence is disformally coupled to relativistic fluid (i.e $\tilde{w} =1/3$ and $\tilde{w}_u =0$). Given the same initial condition of $x=0.01, z= 0.85 , \Sigma =\pi/4)$ on the surface of repeller points, the evolution of the dynamical system  \eqref{E57}, \eqref{E58} and \eqref{E59} depend on the choice of parameters ($\alpha$ and $\beta$). The blue surface represents the boundary of the phase space and satisfies the Friedmann contraint \eqref{E37}, which becomes $3 x^2 \left(\sigma z^2-1\right)+3=4 z^2$. The different trajectories  in the $xz\sigma$ phase space correspond to different combination of parameters. The blue trajectory ($\alpha =0.05$) and green trajectory ($\alpha =1.00$) overlap implying that $\alpha$ does not affect the trajectories on the dynamical system.}
\end{figure} 

 However, the disformal point i.e. $(B_{\pm})_{\textbf{II}}$ is not physical and  the conformal framework ($\Sigma =0$) for this scenario II, does not depend on the parameters $\alpha$ and $\beta$. The $z\sigma$ surface at $x=0$ and $\forall \sigma$, however, consists of a surface of repeller fixed points and any point on that surface should be adequate for analysing how our dynamical system behaves with different combination of parameters. Since an IC which is further away from the equilibrium line $L_{\rm{E}}$ produce appreciable trajectories, we choose the IC as ($x_{\rm{ic}}=0.01,\; z_{\rm{ic}}= 0.85,\; \Sigma_{\rm{ic}} = \pi /4$). The different combination of  parameters ($\alpha,\beta$) can now be tested and we make the following remarks according to figure \ref{fig:ic-sc5}:

\begin{enumerate}
\item The IC are set close to $z\sigma$ with $x\geq 0$ and further away from the equilibrium line $L_{\rm{E}}$, the trajectory gets attracted, as expected nearer to the saddle point $(S_{+})_{\textbf{II}}$ than $A_0$ before ending onto the equilibrium line $L_{\rm{E}}$.

\item The blue trajectory ($\alpha =0.05$) and green trajectory ($\alpha =1.00$) overlap each other which implies that $\alpha$ does not affect the trajectories on the dynamical system \eqref{E57}, \eqref{E58} and \eqref{E59}.

\item The value of $\beta$ determines, where  trajectory ends on the equilibrium line $L_{\rm{E}}$. If the value $\beta$ is large, the trajectory ends near the saddle point $(S_{+})_{\textbf{II}}$ and if the value $\beta$ is small, the trajectory ends near $A_0$. 

\item There is also no effect on the trajectories when the parameter $\alpha \rightarrow -\alpha$.  but if $\beta \rightarrow -\beta$, then the trajectory ends up only at $A_0$, such that they are repelled from $ (S_{+})_{\textbf{II}}$.
\end{enumerate}

\subsection*{8.3.1 Disformal Scenario III: Fixed ($\alpha,\beta$) values but various Initial Conditions}
\vspace*{0.25cm}

\begin{figure}[h!] 
\vspace*{-0.1cm}
\centering
\includegraphics[scale=0.425]{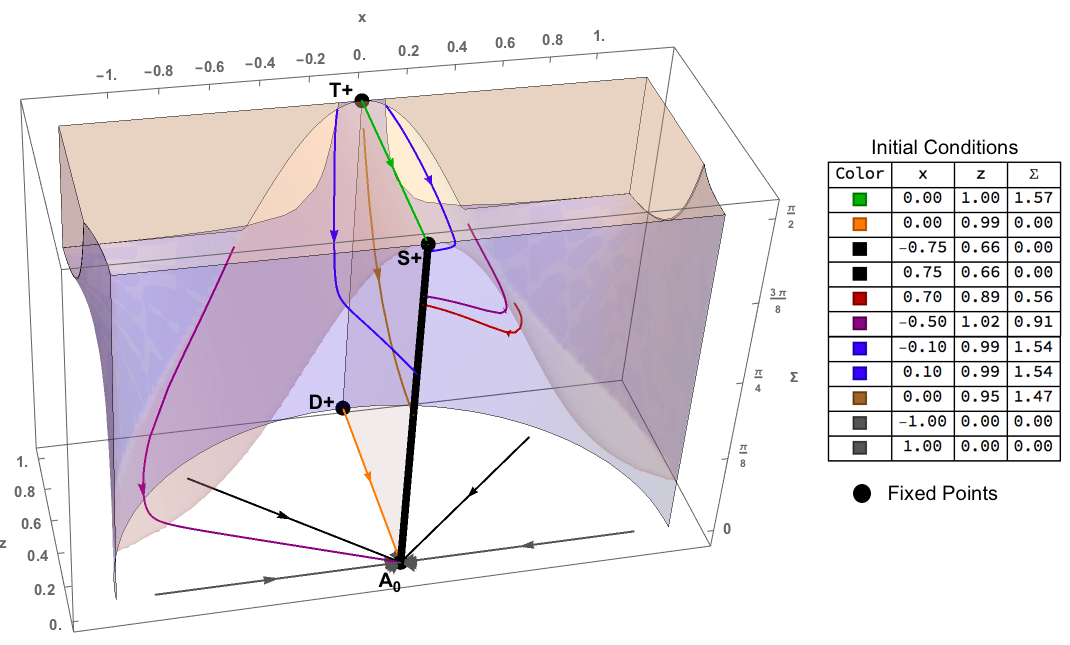} 
\caption{\label{fig:ic-sc6}This figure shows the scenario III, where the quintessence is disformally coupled to relativistic fluid (i.e $\tilde{w} =1/3$ and $\tilde{w}_u =1/3$) with $\alpha = 0.05, \;\text{and}\; \beta = 1.5$  and the $x\sigma z$ phase portrait is shown with the different trajectories, which correspond to different chosen initial conditions. The blue surface represents the boundary of the phase space and satisfies the Friedmann constraint \eqref{E37}, which becomes $4 = 4 z^2 + 3 x^2 (1 - \sigma z^2)$. The black solid line is the equilibrium line $L_{\rm{E}}$ i.e. $(x=0,z=0, \forall\; \sigma)$. The $xz$ plane ($\sigma =0$) is the conformal invariant submanifold and the fixed point $A_0$ is an attracting end point. The $z\sigma$ plane constitutes a topological separation between $x<0$ and $x>0$. The shaded region  is where the condition \eqref{E46} is not satisfied.}
\end{figure}

We now move to the analysis of the trajectories in the scenario III in the next two subsections (i.e $\tilde{w} =1/3$ and $\tilde{w}_u =0$). The values  $\alpha = 0.05,\; \beta = 1.5$ are fixed for sake of comparison. Similar to the previous subsections, few different initial conditions were chosen near the fixed points e.g. $(T_{+})_{\textbf{III}}$ and $(D_{+})_{\textbf{III}}$. Additionally, few other different initial conditions were chosen on edge of the phase portrait on the line of intersection between the boundary surface satisfying the Friedmann constraint i.e. $4 = 4 z^2 + 3 x^2 (1 - \sigma z^2)$ and the region satisfying \eqref{E46}. The reason to choose those initial conditions is to appreciate how the trajectories span out unto the phase portrait. We make the following remarks, according to figure \ref{fig:ic-sc6}:

\begin{enumerate}
\item We recover the conformal behavior (compare with Figure \ref{fig:stream-sc1}) on the plane of $\Sigma = 0$ as expected. The dynamical variable $\Sigma$ gives an additional degree of freedom for the dynamical system to evolve. 

\item When the IC is chosen from the fixed point $(T_{+})_{\textbf{III}}$, the trajectory ends at $(S_{+})_{\textbf{III}}$ and when the IC is chosen from the fixed point $(D_{+})_{\textbf{III}}$, the trajectory ends at $ A_{0}$.

\item All these trajectories whose IC are on the line of intersection between the surface of the blue surface shown as $4 = 4 z^2 + 3 x^2 (1 - \sigma z^2)$ and the surface satisfying \eqref{E46}, flow away from the repelling point  $(T_{+})_{\textbf{III}}$ originally towards an endpoint on the equilibrium line $L_{\rm{E}}$.

\item All trajectories for all the IC without exception end eventually on the equilibrium line $L_{\rm{E}}$. If the IC have $x_{ic} <0$, the trajectories tend to approach the attractor $ A_{0}$ and if the IC have $x_{ic} >0$, the trajectories tend to approach the saddle fixed point $(S_{+})_{\textbf{III}}$.
\end{enumerate}

\subsection*{8.3.2 Disformal Scenario III: Fixed Initial Conditions but various ($\alpha, \beta$) values}
\begin{figure}[h!]
\hspace*{-1.0cm}
\includegraphics[scale=0.425]{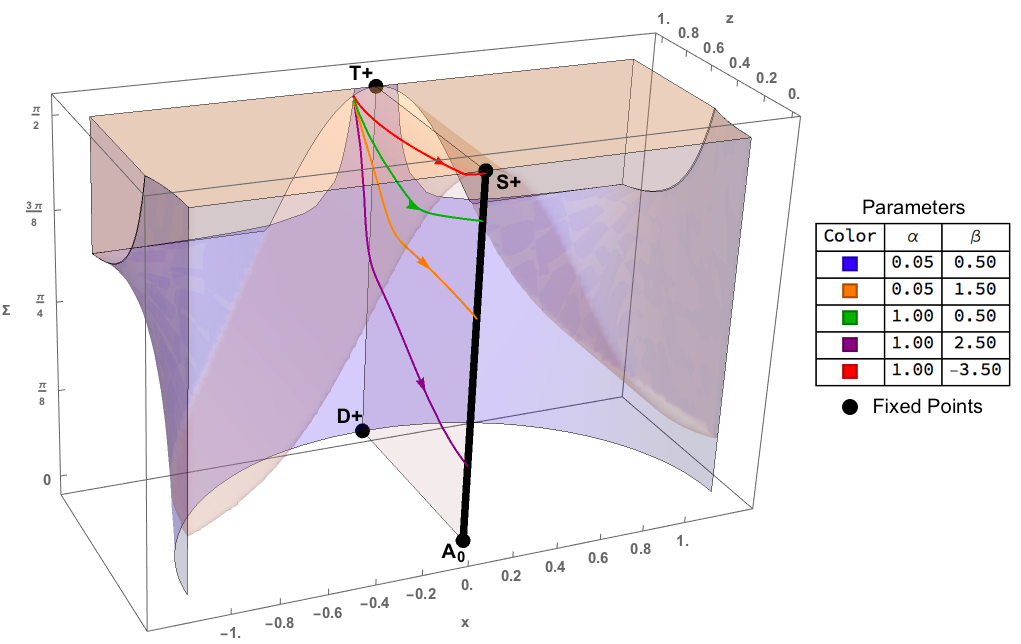} 
\caption{\label{fig:param-sc3}This figure shows the scenario III, where the quintessence is disformally coupled to relativistic fluid (i.e $\tilde{w} =1/3$ and $\tilde{w}_u =1/3$). Given the same initial condition of $(x=-0.1, z= 0.85 , \Sigma =\pi/4)$ on the surface of repeller points, the evolution of the dynamical system \eqref{E64}-\eqref{E67} depends on the parameters $\alpha$ and $\beta$. The blue surface represents the boundary of the phase space and satisfies the Friedmann constraint \eqref{E37}, which becomes $4 = 4 z^2 + 3 x^2 (1 - \sigma z^2)$. The blue trajectory and green trajectory overlap each other which implies that $\alpha$ does not affect the trajectories on the dynamical system. }
\end{figure}

\vspace*{0.25cm}
Similar to the previous subsections, we choose initial conditions (IC) close to a fixed point and evaluate the dependence of the trajectories on the parameters $\alpha$ and $\beta$. The IC was not chosen at $(T_{+})_{\textbf{III}}$ but near it, otherwise the behaviour is assymptotic and regardless of the ($\alpha, \beta$) values, the trajectory flows straight to $(S_{+})_{\textbf{III}}$. But instead, in order to have appreciable effect of the parameters,  the IC are set near the disformal repeller point$(T_{+})_{\textbf{III}}$  i.e. ($x_{ic}=-0.1,z_{ic}=+1.0$, and $\Sigma_{ic} = \pi/2$). The different combination of  parameters ($\alpha,\beta$) can now be tested, and we make the following remarks, given the results of figure \ref{fig:param-sc3}:

\begin{enumerate}
\item The blue trajectory and green trajectory overlap each other, which implies that $\alpha$ does not affect the trajectories on the dynamical system \eqref{E64}-\eqref{E67}.

\item The value of $\beta$ determines, where  trajectory ends on the equilibrium line $L_{\rm{E}}$. If the value $\beta$ is large, the trajectory ends further away from the saddle point $(S_{+})_{\textbf{III}}$ and if the value $\beta$ is small, the trajectory ends close to $(S_{+})_{\textbf{III}}$

\item There is also no effect on the trajectories when the parameter $\alpha \rightarrow -\alpha$.  but if $\beta \rightarrow -\beta$, then the trajectory ends up only at $ (S_{+})_{\textbf{III}}$.
\end{enumerate}

\subsection*{8.4 Summary of all disformal fixed points}

In this subsection, we summarise all disformal fixed points and their eigenvalues which were obtained from the three studied scenarios in the following tables. The conformal fixed points are a subset of the disformal fixed points, irrespective of the scenario.

%&&&&&&&&&&&&&&&&&&&&&&&&&&&&&&&&&&&&&&&&&&&&
\begin{table}[ht]
\begin{tabular}{|| c| c | c | c | c c c | c |c||}
 \hline
 Point & $x$ & $z$ &$\sigma$ & $ \Omega_{\phi} $&  $ w_{\phi} $ &  $w_{\text{eff}}$ &Existence & Acceleration\\ [0.5ex] 
 \hline\hline
$A_{0}$  &$0$& 0 & 0  &1& -1 & 0 &$\forall \beta,\alpha$& Yes\\ 
\hline
$(C_{\pm})_{\textbf{I}}$  & $\pm 1$& 0 & 0 &1& 0& 0 &$\forall \beta,\alpha$& No\\ 
\hline
$(B_{\pm})_{\textbf{I}}$ & $\dfrac{\sqrt{3}}{2 \beta }$ & $\pm\dfrac{\sqrt{4 \beta^2-3}}{2 \beta}$ & $\dfrac{4 \beta ^2}{9}$ &$\dfrac{3}{4 \beta ^2}$& 0 & $\frac{3}{4 \beta ^2}-1$& $\beta \leq -\frac{\sqrt{3}}{2}$ or $\beta \geq \frac{\sqrt{3}}{2}$ & No \\  [1.5ex]  
\hline
 $(S_{\pm})_{\textbf{I}}$ & 0 & 0 &$\pm \infty$ & $ 1 $&  $ - $ &  $-$ & $\forall \beta,\alpha$& -\\
\hline
 $(T_{\pm})_{\textbf{I}}$ & 0 & $\pm 1$ &$\pm \infty$ & $ 0 $&  $ - $ &  $-$ & $\forall \beta,\alpha$& -\\
\hline
\end{tabular}
\caption{\label{table:points-sc1} This table shows a summary of all the fixed points which was obtained by solving the dynamical system \eqref{E49}, \eqref{E50} and \eqref{E51} for a quintessence  disformally coupled with pressureless fluid in the Scenario I ($\tilde{w} =0$ and $\tilde{w}_u =0$). For each fixed point, the existence and the effective equation of state, as defined in \eqref{E43}, are evaluated. The $``-"$ means undetermined.}
\end{table}
%&&&&&&&&&&&&&&&&&&&&&&&&&&&&&&&&&&&&&&&&&&&&
\begin{table}[ht]
 \begin{tabular}{||  l |  c  c   c |}
 \hline
& Conformal &&\\
 \hline
Point & $E_{1}$&$E_{2}$ &\hspace*{0.5cm}$E_{3}$\hspace*{0.5cm}\\
 \hline
Point $A_0$ &$-\frac{3}{2}$&$-\frac{3}{2}$ &$0$ \\[1.2ex]  
 \hline 
 Point $(C_{\pm})_{\textbf{I}}$ &$+3$& $+\dfrac{3}{2}$ & $\pm\frac{\sqrt{3} \alpha }{2}$ \\[1.2ex]  
 \hline 
% && &\\[1.2ex]  
 %\hline 
\end{tabular}
 \begin{tabular}{||  l |  c  c   c |}
 \hline
& Disformal &&\\
 \hline
Point & $E_{1}$&$E_{2}$ &\hspace*{0.cm}$E_{3}$\hspace*{0.7cm} \\
 \hline
Point $A_0$ &$-\frac{3}{2}$&$-\frac{3}{2}$ &$0$ \\[1.2ex]  
 \hline 
 Point $(C_{\pm})_{\textbf{I}}$ &$+3$& $\pm \frac{\sqrt{3} \alpha }{2}$ & $\pm 2 \sqrt{3} \beta -3$ \\[1.2ex]  
 \hline 
 % Point $(B_{\pm})_{\textbf{I}}$ &$\frac{3 \alpha }{\beta }$& $-\frac{3}{2}$ & $3$ \\[1.2ex]  
 %\hline 
\end{tabular}
\caption{\label{table:eigen-sc1} This table shows all the eigenvalues, which are evaluated at each fixed point, for scenario I ($\tilde{w} =0$ and $\tilde{w}_u =0$). The left tabular corresponds to the eigenvalues of the conformal dynamical system \eqref{E47} and  \eqref{E48}. The right tabular corresponds to the eigenvalues of the disformal dynamical system \eqref{E49}, \eqref{E50} and \eqref{E51}. The eigenvalues of the disformal fixed points $(B_{\pm})_{\textbf{I}}$, $(S_{\pm})_{\textbf{I}}$ and $(T_{\pm})_{\textbf{I}}$ are undetermined.}
\end{table}
%&&&&&&&&&&&&&&&&&&&&&&&&&&&&&&&&&&&&&&&&&&&&
\begin{table}[ht]
 \begin{tabular}{|| c | c | c |  c |  c |  c |  c | c |c||}
 \hline
 Point &$x$&$z$&$\sigma$ & $ \Omega_{\phi} $&  $ w_{\phi} $ &  $w_{\text{eff}}$ &Existence & Acceleration \\
 \hline\hline
$ A_{0}$  &$0$& 0 & 0 &1& -1 &-1&$\forall \beta,\alpha$& Yes \\ 
\hline
$(C_{\pm})_{\textbf{II}}$  & $\pm 1$& 0 & 0 &1& 0 &0&$\forall \beta,\alpha$ & No\\ 
\hline
$(B_{\pm})_{\textbf{II}}$ & $\dfrac{\sqrt{3}}{2 \beta }$ & $\pm\dfrac{\sqrt{4 \beta^2-3}}{2 \beta}$ & $\dfrac{4 \beta ^2}{9}$ &$\dfrac{3}{4 \beta ^2}$& 0 & $\frac{3}{4 \beta ^2}-1$& $\beta \leq -\frac{\sqrt{3}}{2}$ or $\beta \geq \frac{\sqrt{3}}{2}$ & No \\  [1.5ex]  
\hline
 $(S_{\pm})_{\textbf{II}}$ & 0 & 0 &$\pm \infty$ & $ 1 $&  $ - $ &  $-$ & $\forall \beta,\alpha$& -\\
\hline
\end{tabular}
\caption{\label{table:points-sc2}A summary of all the fixed points for the dynamical system \eqref{E57}-\eqref{E59} for a quintessence  disformally coupled with relativistic fluid in the Scenario II ($\tilde{w} =1/3$ and $\tilde{w}_u =0$).  The $``-"$ means undetermined.}
\end{table}
%&&&&&&&&&&&&&&&&&&&&&&&&&&&&&&&&&&&&&&&&&&&&
\begin{table}[ht]
 \begin{tabular}{||  l |  c  c   c |}
 \hline
& Conformal &&\\
 \hline
Point & $E_{1}$&$E_{2}$ &\hspace*{0.5cm}$E_{3}$\hspace*{0.5cm} \\
\hline
Point $A_0$ &$-2$&$-\frac{3}{2}$ &$0$ \\[1.4ex]  
 \hline 
Point $(C_{\pm})_{\textbf{II}}$ &$+3$& $+\dfrac{3}{2}$ & $-\dfrac{1}{2}$ \\[1.4ex]    
 \hline
&&&\\[1.2ex]   
 \hline 
\end{tabular}
 \begin{tabular}{||  l |  c  c   c |}
 \hline
& Disformal &&\\
 \hline
Point & $E_{1}$&$E_{2}$ &\hspace*{0.7cm}$E_{3}$\hspace*{0.7cm} \\
 \hline
Point $A_0$ &$-2$&$-\frac{3}{2}$ &$0$ \\[1.2ex]   
 \hline 
 Point $(C_{\pm})_{\textbf{II}}$ &$+3$& $-\dfrac{1}{2}$ & $\pm 2 \sqrt{3} \beta -3$ \\[1.2ex]  
 \hline 
  Point $(B_{\pm})_{\textbf{II}}$ &$-2$& $-\dfrac{3}{2}$ & $+ 3$ \\[1.2ex]  
 \hline 
\end{tabular}
\caption{\label{table:eigen-sc2}This table shows all the eigenvalues, which are evaluated at each fixed point, for scenario II ($\tilde{w} =1/3$ and $\tilde{w}_u =0$). The left tabular corresponds to the eigenvalues of the conformal dynamical system \eqref{E57S3} and  \eqref{E58S3}. The right tabular corresponds to the eigenvalues of the disformal dynamical system \eqref{E57}, \eqref{E58} and \eqref{E59}. The disformal eigenvalues of $(S_{\pm})_{\textbf{II}}$ are undertermined.}
\end{table}
%&&&&&&&&&&&&&&&&&&&&&&&&&&&&&&&&&&&&&&&&&&&&
\begin{table}[ht]
 \begin{tabular}{|| c | c | c |  c |  c |  c |  c | c |c|c|c|c||}
 \hline
 Point &$x$&$z$&$\sigma$ & $ \Omega_{\phi} $&  $ w_{\phi} $ &  $w_{\text{eff}}$ &Existence & Acceleration & $E_{1}$&$E_{2}$ &$E_{3}$ \\
 \hline\hline
$ A_{0}$  &$0$& 0 & 0 &1& -1 &-1&$\forall \beta,\alpha$& Yes &$-2$&$-2$ &$0$ \\ 
\hline
$(D_{\pm})_{\textbf{III}}$  & 0 & $\pm 1$& 0 &0& - &$\dfrac{1}{3}$&$\forall \beta,\alpha$& No &$-4$&$-4$ &$0$\\ 
\hline
 $(S_{\pm})_{\textbf{III}}$ & 0 & 0 &$\pm \infty$ & $ 1 $&  $ - $ &  $-$ & $\forall \beta,\alpha$& -&$- $&$- $&$- $\\
\hline
 $(T_{\pm})_{\textbf{III}}$ & 0 & $\pm 1$ &$\pm \infty$ & $ 0 $&  $ - $ &  $-$ &  $\forall \beta,\alpha$ & -&$- $&$- $&$- $\\
\hline
\end{tabular}
\caption{\label{table:points-sc3}This table shows a summary of all the fixed points for the Scenario III ($\tilde{w} =1/3$ and $\tilde{w}_u =1/3$). The $``-"$ means undetermined. This table also includes the Eigenvalues which are evaluated at these fixed points for the disformal dynamical system  \eqref{E64}-\eqref{E67}. The fixed point $A_0$ has same Eigenvalues for the conformal dynamical system \eqref{E62b}-\eqref{E63}. The nature of fixed point $A_0$ and $(D_{\pm})_{\textbf{III}}$  are found to be stable and unstable respectively using Lyapunov theorem. }
\end{table}

\newpage
\pagebreak
\section{Cosmological Analysis}
%%%%%%%%%%%%%%%%%%%%%%%%%%%%%%%%%%%%%%%%%%%%%%%%%%
After performing an analysis of the dynamical system for each disformal scenario, we will consider a universe with a radiation-dominated phase, and DM dominated phase, before it enters a de-Sitter evolution. In this section, the expansion history of this coupled quintessence model (which is constructed to mimic $\Lambda$CDM background)  is studied in terms of cosmological redshift $Z = 1/a -1 $. The redshift $Z$ is not to be confused with dynamical variable $z$. The quintessence is coupled to DM throughout the cosmological timeline, implying that $z$  represents the energy  density of DM, therefore this is expansion history of the cosmological model in disformal scenario I, whose dynamical system is described by \eqref{E49}-\eqref{EQ1}
\begin{figure}[h!] 
\centering
\includegraphics[scale=0.4]{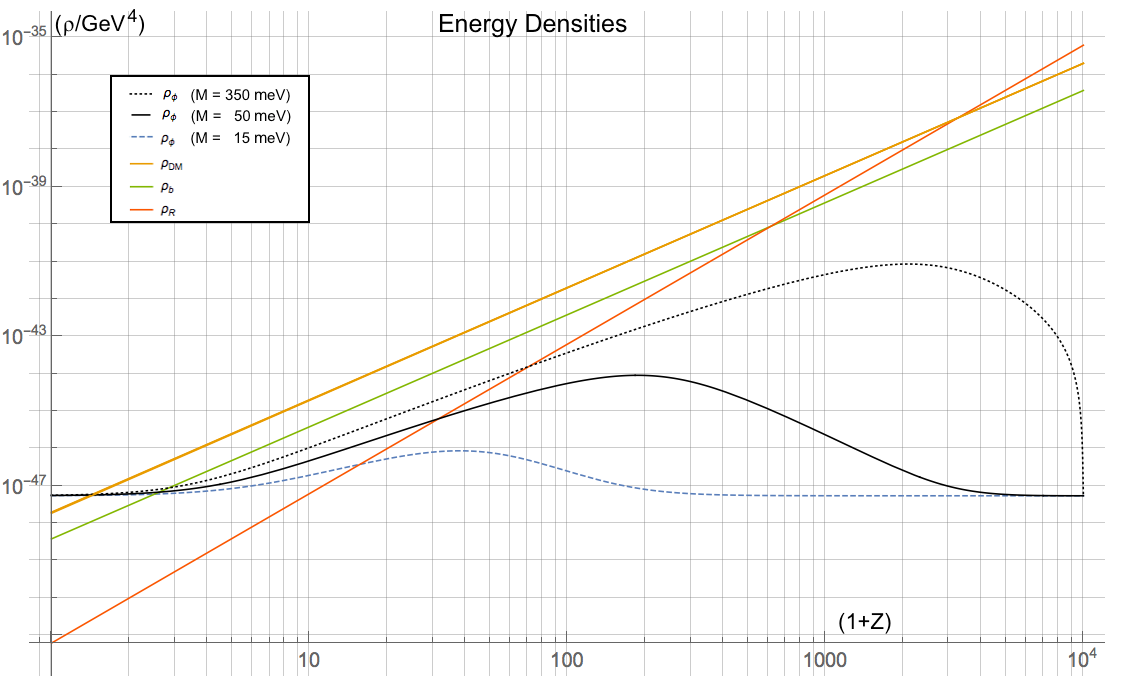} 
\caption{\label{fig:energies}The energy density for  the quintessence, radiation, DM and baryons, obtained when the dynamical system is evolved from the early universe ($Z_i=10^4$) to present using three different mass scales of the quintessence.}
\end{figure}

Contrary to the three previous sections, where the initial values were theoretical and motivated by the fixed points on the phase space, it also makes sense to consider more realistic initial conditions for our universe, similar to $\Lambda$CDM background at redshift $Z_i = 10^4$. \\

From the above definition \eqref{E36}, one evaluates the initial values of the dynamical variables as well at redshift $Z_{i}$ for an early universe according to the relations,
\begin{eqnarray}  \label{E68}
x_{i}=0 ,\hspace{0.5cm}
y_{i}=\sqrt{\frac{\Omega _{\Lambda o}}{h^2_{i}}},\hspace{0.5cm}
z_{i}&=&\sqrt{\frac{\Omega _{\rm{cdm}0}(1+Z_{i})^3}{h^2_{i}}},\hspace{0.5cm} 
\sigma_{i} =  \dfrac{D H^2_{0}}{\kappa^2 C}h^2_{i}, \hspace{0.5cm}
h_{i}=h(Z_i),
\end{eqnarray}
%h^2(r) = \Omega _{\Lambda o} +\Omega _{bo}(1+r)^3+ \Omega _{cdmo}(1+r)^3+ \Omega _{\gamma o}(1+r)^4
where $h^2(Z) \equiv H^2/H_{0}^2$, is the reduced Hubble function and $H_{0}$ is Hubble's constant.The density parameters $\Omega _{\rm{cdm}0}$ and $ \Omega _{\Lambda 0}$ correspond to DM and DE evaluated today.  

The parameters are set to $\alpha = 0.05$ and $ \beta = 1.5$, which are chosen similar to the two previous sections and the value of $x_{i}$  is set to zero, as suggested by \cite{Barros2018}. Additionally from equation \eqref{E34}, it can be noticed that $D(\phi_{i}) \propto M^{-4}$, thus the mass scale of the coupled quintessence becomes relevant in specifying the initial value for $\sigma_{i}$ in \eqref{E68}. In the literature \citep{Carsten2013,vandeBruck2015c}, it has been constrained that the mass scale $M$ is at least larger than $M_{1} \geq 15\; \rm{meV}$. In addition to this masses $M_{1}$, we consider two hypothetical mass scales ($M_{2} = 50 \; \rm{meV},\; \rm{and}\;$ $M_{3} = 350 \; \rm{meV}$) to test how large $M$ must be in order to render the disformal contribution negligible. We hence investigate how the different mass scales $M_{1},\; M_{2}$ and $M_{3}$ influence the dynamical system \eqref{E49}-\eqref{EQ1}. Figure \ref{fig:energies} shows the expansion history of this coupled quintessence model in terms of cosmological redshift and how the energy densities of the quintessence, radiation, DM and baryons evolves  from $Z_i = 10^4$ to the present. 
%%%%%%%%%%%%%%%%%%%%%%%%%%%%%%%%%%%%%%%%%
The following remarks were noted based on the results from figure \ref{fig:energies}:
\begin{enumerate}
\item The features of $\Lambda$CDM are reproduced as expected such as the radiation-matter transition and the DM-DE transition. 

\item The effect of different mass scales clearly affects the evolution of the energy density for quintessence $\rho_{\phi}$ as seen in figure \ref{fig:energies} and its density parameter $\Omega_{\phi} = x^2 + y^2$. However, from definition \eqref{E36}, the dynamical variables $x$ and $y$ do not depend on the mass. Rather, this effect only happens because of $\sigma$, which measures the strength of the disformal coupling, and depends on the mass $M$, given the definition of $D(\phi)$ in \eqref{E34}. This effect shows the deviation from the conformal framework ($\sigma =0 $), which is more apparent and important at early times.
\item The energy density of the quintessence traces that of a cosmological constant, when low mass scale is utilised. A larger mass scale causes deviation from the conformal framework.
\end{enumerate}

\begin{figure}[h!] 
\centering 
\includegraphics[scale=0.375]{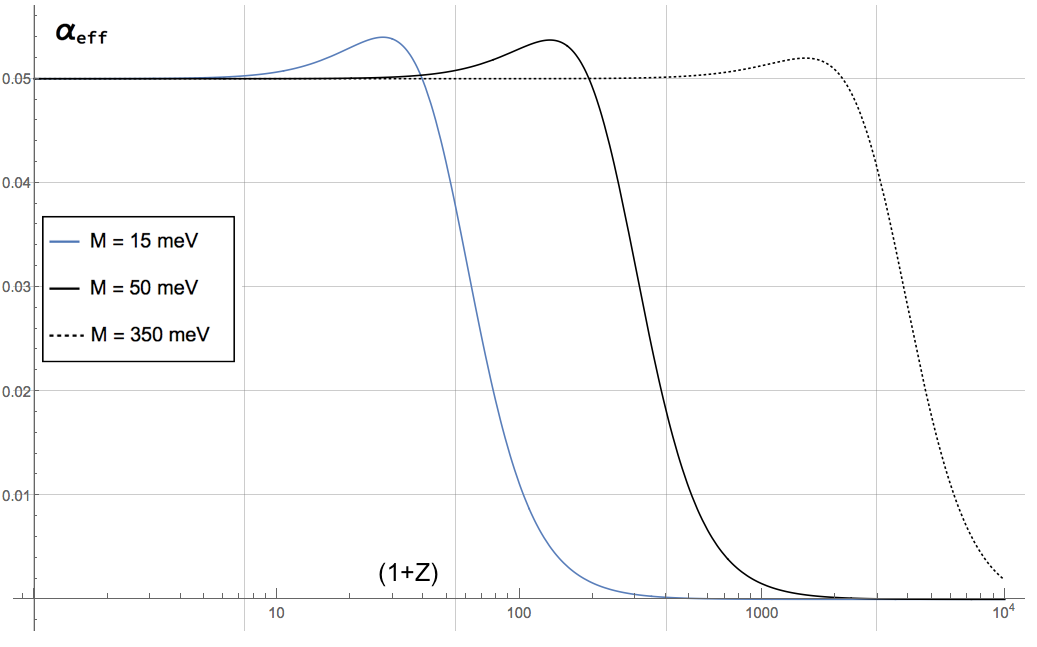} 
\caption{\label{fig:alphaeff}The evolution of the effective conformal coupling ($\alpha_{\text{eff}}$) from initial redshift $Z_i = 10^4$ to present was carried out for three mass scales $M_{1}, M_{2}$ and $M_{3}$. The turn over point, i.e. when the bump occurs, indicates the redshift when the contribution of the disformal coupling is of the same magnitude as the conformal coupling. After the turn over point, the effective conformal coupling $\alpha_{\text{eff}} \to \alpha$, because the disformal coupling becomes insignificant.}
\end{figure}

Now that we understand that the disformal framework plays a more important role at early times, it becomes relevant to investigate how the disformal nature of the dynamical system \eqref{E49}-\eqref{EQ1} evolves from the early times to present. As mentioned in Section VII.1., in a purely conformal framework, the interaction term $Q$ does reduce to $Q =  \alpha \rho\nabla^{b}\phi$ \cite{Amendola1999}, which is function of $\alpha$ only. In essence, the general form of $Q(\phi)$ defined in \eqref{E32} which is a function of $\alpha$ and $\beta$, behold both the disformal and conformal character of the dynamical system. It makes sense to define an effective conformal coupling \cite{Teixeira2019}:
\begin{equation}
\alpha_{\text{eff}} = -\dfrac{Q}{\kappa \rho}\;,
\end{equation}
Where $Q(\phi)$ is initially defined as in equation \eqref{E40} but reduces to  equation \eqref{EQ1} in this scenario and $\rho$ is related to dynamical variable $z$, as in \eqref{E36}. In other words, if $\alpha$ characterises only the rescaling of the metric via the conformal transformation  \eqref{E1}, and $\beta$ characterises only the deviation from pure conformal framework, i.e., distortion of angles of the metric by the disformal transformation \eqref{E2}, $\alpha_{\text{eff}}$ provides an interplay of both these characteristics. Figure \ref{fig:alphaeff} shows the evolution of the $\alpha_{\text{eff}}$ with redshift $Z$ for the three different mass scales $M_{1}, M_{2}$ and $M_{3}$. From the results of figure \ref{fig:alphaeff}, the following interpretation was noted:
\begin{enumerate}

\item When the $\alpha_{\text{eff}}$ is close to zero, the disformal coupling is cancelling the conformal coupling as shown in \cite{vandeBruck2015}.

\item There exists a turn over point, which indicates the redshift at which the contribution from the disformal coupling is of the same magnitude as the conformal coupling.

\item After the turn over point, the decay of the disformal coupling continues until it becomes insignificant and the $\alpha_{\text{eff}}$ becomes the chosen value of the conformal coupling constant $\alpha = 0.05$. The framework can be approximated as nearly conformal and the dynamical system \eqref{E49}-\eqref{EQ1} reduces to a system described by \eqref{E47} and \eqref{E48} at late times.

\item The $\alpha_{\text{eff}}$ is sensitive to the mass scale, which alters the redshift of the turnover point at which  the disformal coupling becomes insignificant. A heavier (lower) mass scale would cause turnover to occur  at a much higher (lower) redshift.
\end{enumerate}

\begin{figure}[h!]  
\includegraphics[scale=0.39]{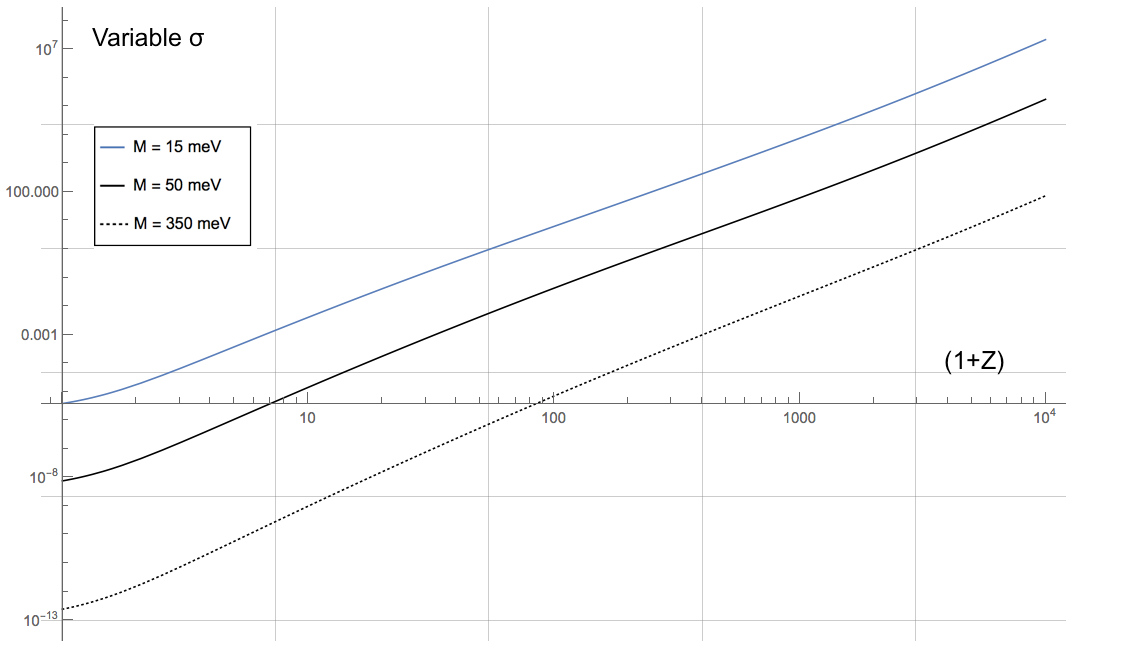} 
\caption{\label{fig:sigma}The decay of the dynamical variable $\sigma$, which  represent the strength of the disformal coupling, is shown to occur at approximately a constant logarithmic rate throughout the whole expansion history for the three mass scales.}
\end{figure}

The dynamical variable $\sigma$, which  represents the strength of the disformal coupling, starts with large value initially but then decays at constant logarithmic rate throughout the whole expansion history  as seen in Figure \ref{fig:sigma}. From definition \eqref{E34} and taking the logarithm of  $\sigma$ in \eqref{E36}, one can find that the mass scale does not affect the rate of decay but only the time when $\sigma$ becomes ${\cal O}(1)$. 

In short, after making the cosmological analysis for coupled quintessence mimicking $\Lambda$CDM,  we find that different mass scales clearly do affect the evolution of the energy density for quintessence $\rho_{\phi}$ in a similar way to what was found in previous studies with an exponential potential \citep{vandeBruck2015,Teixeira2019}. More specifically, the presence of a disformal coupling has the effect of screening the interaction between matter and the scalar field at high redshifts. The time of crossover to the dominant conformal coupling might be revealed when studying the evolution of density perturbations \citep{vandeBruck2015}.

\newpage
\section{Conclusion}
%%%%%%%%%%%%%%%%%%%%%%%%%%%%%%%%%%%%%%%%%%%%%%%%%%
In this article, we investigated the cosmological dynamics of coupled quintessence (where the quintessence $\phi$ interacts with a generic fluid $\rho_c$) in a disformal framework. The novelty of this work is that it extends the idea of coupled quintessence, mimicking the cosmological constant $\rho_{\Lambda}$ and an uncoupled fluid $\rho_u$ \cite{Barros2018}, to include a disformal coupling, and (ii) to bring these conformal and disformal couplings with $\Lambda$CDM background to a dynamical system analysis for the first time in the literature.  This brings the convenience of not having to specify the scalar field potential. 

The choice for EOS parameters in the disformal frame for the coupled fluid and uncoupled fluid i.e. $\tilde{w}$ and $\tilde{w}_u$ respectively give rise to different scenarios. There are three  studied scenarios, which are (a) Scenarios I ($\tilde{w} =0$ and $\tilde{w}_u =0$) consists of a pressureless fluid (e.g DM) is coupled to quintessence and both fluids are mimicking a $\Lambda \rm{CDM}$ background with a cosmological constant $\Lambda$ and CDM, (b) Scenarios II ($\tilde{w} =1/3$ and $\tilde{w}_u =0$) consists of a relativistic fluid is coupled to quintessence and both fluids are mimicking a $\Lambda \rm{CDM}$ background with a cosmological constant $\Lambda$ and CDM, (c) Scenarios III ($\tilde{w} =1/3$ and $\tilde{w}_u =1/3$) consists of a relativistic fluid coupled to quintessence and both fluids are mimicking a background with a cosmological constant $\Lambda$ and an uncoupled relativistic fluid. Section VI entails the analysis of the conformal couplings in the three studied scenarios, which includes the conformal equations, the fixed points, and 2D phase portrait. Section VII entails the analysis of the disformal couplings in the three studied scenarios, which includes the disformal equations, the additional disformal fixed points, the topological features and the investigation of the  trajectories on a 3D phase portrait.

\newpage
The differential equations \eqref{E38}-\eqref{E42} which govern the behaviour for the dynamical system bring novel phenomenology in each scenario regarding particular solutions (fixed points) and dynamics connecting such solutions. The nature of the fixed points were studied and their conditions for them to be either stable, unstable and saddle were given. It is worth briefly mentioning a few comparisons for which we find the following differences. In the scenarios I , II, and III, there are nine, seven and seven fixed points respectively. The conformal framework for  scenario II and scenario III, does not depend on the parameters $\alpha$ and $\beta$. The unstable point $(D_{+})_{\textbf{II}}$ is radiation-dominated point which appears only in the scenario III. The repeller point $(T_{+})_{\textbf{II}}$ does not appear in the scenario II. There exists a separating topological feature, namely $z\sigma$ plane, in scenario II and III but is absent in scenario I. The physical fixed points in scenario I are $A_0, (C_{\pm})_{\textbf{I}}, (S_{+})_{\textbf{I}} , (T_{+})_{\textbf{I}}  $, and in scenario II are $A_0, (C_{\pm})_{\textbf{II}},(S_{+})_{\textbf{II}} $ and lastly in scenario III are $A_0, (D_{+})_{\textbf{III}}, (S_{+})_{\textbf{III}} , (T_{+})_{\textbf{III}} $. In all scenarios, we commonly find the invariant sub-manifold $xz$ plane, the attractor $A_0$, $S_{\pm}$, and the equilibrium line $L_{\rm{E}}$ i.e $(x=0,z=0, \forall\; \sigma)$, as well as the shaded region in which \eqref{E46} is not satisfied.  A rich phenomenology of the dynamical system  can be understood when comparing between scenarios. 

  Further analysis focused more on the trajectories for fixed initial conditions and various parameters ($\alpha$ and $\beta$). For each scenario, we recover the conformal behaviour on the plane of $\Sigma = 0$ as expected (Compare top panel of Figure \ref{fig:stream-sc1} with Figure \ref{fig:ic-sc1}, and similarly bottom left panel of Figure \ref{fig:stream-sc1} with Figure \ref{fig:ic-sc2}, and lastly bottom right panel of Figure \ref{fig:stream-sc1} with Figure \ref{fig:ic-sc6}). The dynamical variable $\Sigma$ gives an additional degree of freedom for the dynamical system to evolve. For instance in all scenarios, there is an equilibrium line of fixed points, namely $L_{\rm{E}}$. Particularly, in scenario II, there is an entire plane of repeller fixed points to consider. For each scenario, all trajectories end on the equilibrium line  $L_{\rm{E}}$, which connects the attractor $A_0$ and $S_{+}$. The dynamical system flows from one fixed point to another within this closed compactified space. The dynamical system is sensitive to the chosen initial conditions. i.e. different initial conditions around the proximity of the fixed points can lead to different endpoints. For instance in scenario I, when the IC is chosen near $(C_{+})_{\textbf{I}}$, the dynamical system can end up at either $A_{0}$ or on the equilibrium line near$(S_{+})_{\textbf{I}}$. Similarly, IC is chosen near $(T_{+})_{\textbf{I}}$, the endpoints  could be either at $A_{0}$ or $(S_{+})_{\textbf{I}}$ or on the equilibrium line $L_{\rm{E}}$. The analysis of the trajectories was also done for fixed initial conditions but different parameters ($\alpha$ and $\beta$). One remark from scenario I is that the parameters ($\alpha,\beta$) determines how much the dynamical system  \eqref{E49}, \eqref{E50} and \eqref{E51} gets influenced by either the conformal fixed points $(C_{\pm})_{\textbf{I}}$ or the disformal fixed point $(S_{+})_{\textbf{I}}$, as it evolves on its trajectory within the phase space. Another remark from scenario II and III, that $\alpha$ does not affect the trajectories on the dynamical system. Furthermore in scenario II and III, the value of $\beta$ determines, where  trajectory ends on the equilibrium line $L_{\rm{E}}$. The two analysis of the trajectories allowed us to better understand how the choice of IC and parameters affect the dynamical system.
  
 A more realistic cosmological analysis of the coupled quintessence was then carried out where the quintessence is coupled to DM throughout the cosmological evolution. Therefore this expansion history corresponds to the cosmological model in disformal scenario I, whose dynamical system is described by \eqref{E49}-\eqref{EQ1}. The features of $\Lambda$CDM are reproduced as expected such as the radiation-matter transition and the DM-DE transition. By construction, our dynamical system is exactly the same as $\Lambda$CDM at late-time cosmologies. The effect of different mass scales clearly affects the evolution of the energy density for quintessence $\rho_{\phi}$ as seen in figure \ref{fig:energies} and its density parameter $\Omega_{\phi}$. This effect shows the deviation from the conformal framework ($\sigma =0 $), which is more important at early times. The energy density of the quintessence is effectively mimicking a cosmological constant, when low mass scale is utilised. A larger mass scale will cause greater deviation from the conformal framework. The effect of the mass scale of the quintessence influences the disformal characteristics of the dynamical system, which is quantified by the effective conformal coupling (see Figure \ref{fig:alphaeff}). This behaviour is similar to what was found in \cite{Teixeira2019}. There exists a turn over point, which indicates the redshift at which the contribution from the disformal coupling is of the same magnitude as the conformal coupling. After the turn over point, the decay of the disformal coupling continues until it becomes insignificant and the $\alpha_{\text{eff}}$ becomes the chosen value of the conformal coupling constant. The $\alpha_{\text{eff}}$ is sensitive to the mass scale, which alters the redshift of the turnover point at which  the disformal coupling becomes insignificant. A heavier (lower) mass scale would cause turnover point at much higher (lower) redshift. 
 
It would be interesting to further explore this model by studying the evolution of linear density perturbations and test them against Redshift Space Distortion data. It is also important to study the spherical collapse of over-densities when going to the non-linear evolution.

\acknowledgments
AdlCD and AD acknowledge financial support from NRF Grants No.120390,
Reference: BSFP190-416431035, and No.120396, Reference: CSRP190405427545, and No 101775, Reference: SFH1507-27131568. AdlCD also acknowledges financial support from Project No. FPA2014-53375-C2-1-P from the Spanish Ministry of Economy and Science, MICINN Project No. PID2019-108655GB-I00, Project No. FIS2016-78859-P from the European Regional Development Fund and Spanish Research Agency (AEI), and support from Projects Nos. CA15117 and CA16104 from COST Action EU Framework Programme Horizon 2020. AdlCD \& PD thanks the hospitality of the Institute of Theoretical Astrophysics - University of Oslo (Norway) during the later steps of the manuscript. AdlCD and NN also acknowledge funding from the University of Cape Town Visiting Scholars Fund 2018.  
This research was supported by Funda\c{c}\~ao para a Ci\^encia e a Tecnologia (FCT) through the research grants: UID/FIS/04434/2019, PTDC/FIS-OUT/29048/2017 (DarkRipple),  COMPETE2020: POCI-01-0145-FEDER-028987 \& FCT: PTDC/FIS-AST/28987/2017 (CosmoESPRESSO), CERN/FIS-PAR/0037/2019 (MGiCAP) and IF/00852/2015 (Dark Couplings). 

%\newpage

\end{document}